\documentclass[english, superscriptaddress, preprint, aps, amsmath,amssymb]{revtex4-1}
\usepackage[latin9]{inputenc}
\usepackage{graphicx}
\usepackage{xcolor}
\newcommand{\Spi}{Sierpi\'{n}ski }
\usepackage{amsmath}
\usepackage[resetlabels, labeled]{multibib}
\setcitestyle{super}
\newcites{S}{References Supplementary Materials}

\usepackage{tikz}
\usepackage{tocvsec2}
\usepackage{scrextend}
\usepackage{lineno}
\modulolinenumbers[5]
\makeatletter
\@ifundefined{textcolor}{}
{
 \definecolor{BLACK}{gray}{0}
 \definecolor{WHITE}{gray}{1}
 \definecolor{RED}{rgb}{1,0,0}
 \definecolor{GREEN}{rgb}{0,1,0}
 \definecolor{BLUE}{rgb}{0,0,1}
 \definecolor{CYAN}{cmyk}{1,0,0,0}
 \definecolor{MAGENTA}{cmyk}{0,1,0,0}
 \definecolor{YELLOW}{cmyk}{0,0,1,0}
}

\makeatother

\usepackage{babel}

\begin{document}
\title{Design and characterization of electrons in a fractal geometry}
\author{S. N. Kempkes}
\thanks{Both authors contributed equally.}
\affiliation{Institute for Theoretical Physics, Utrecht University, Netherlands}
\author{M.R. Slot}
\thanks{Both authors contributed equally.}
\affiliation{Debye Institute for Nanomaterials Science, Utrecht University, Netherlands}
\author{S. E. Freeney}
\affiliation{Debye Institute for Nanomaterials Science, Utrecht University, Netherlands}
\author{S. J. M. Zevenhuizen}
\affiliation{Debye Institute for Nanomaterials Science, Utrecht University, Netherlands}
\author{D. Vanmaekelbergh}
\affiliation{Debye Institute for Nanomaterials Science, Utrecht University, Netherlands}
\author{I. Swart}
\email{Correspondence to: C.deMoraisSmith@uu.nl, I.Swart@uu.nl}
\affiliation{Debye Institute for Nanomaterials Science, Utrecht University, Netherlands}
\author{C. Morais Smith}
\email{Correspondence to: C.deMoraisSmith@uu.nl, I.Swart@uu.nl}
\affiliation{Institute for Theoretical Physics, Utrecht University, Netherlands}

\date{\today}
\maketitle
\linenumbers

\textbf{The dimensionality of an electronic quantum system is decisive for its properties. In 1D electrons form a Luttinger liquid and in 2D they exhibit the quantum Hall effect. However, very little is known about the behavior of electrons in non-integer, i.e. fractional dimensions~\cite{Mandelbrot}. Here, we show how arrays of artificial atoms can be defined by controlled positioning of CO molecules on a Cu(111) surface~\cite{Manoharan,Slot2017,Collins2017}, and how these sites couple to form electronic \Spi fractals. We characterize the electron wavefunctions at different energies with scanning tunneling microscopy and spectroscopy and show that they inherit the fractional dimension. Wavefunctions delocalized over the \Spi structure decompose into self-similar parts at higher energy, and this scale invariance can also be retrieved in reciprocal space. Our results show that electronic quantum fractals can be man-made by atomic manipulation in a scanning tunneling microscope. The same methodology will allow to address fundamental questions on the effects of spin-orbit interaction and a magnetic field on electrons in non-integer dimensions. Moreover, the rational concept of artificial atoms can readily be transferred to planar semiconductor electronics, allowing for the exploration of electrons in a well-defined fractal geometry, including interactions and external fields.}

Fractals have been investigated in a wide variety of research areas, ranging from polymers~\cite{Newkome}, porous systems~\cite{Yu}, electrical storage~\cite{Dubal} and stretchable electronics~\cite{Fan} down to molecular\cite{Newkome, Rothemund, Shang, Li} and plasmonic\cite{Nicola} fractals.  
On the quantum level, fractality emerges in the behavior of electrons under perpendicular magnetic fields (Hofstadter butterfly~\cite{Hofstadter}, quantum Hall resistivity curve~\cite{Pan, Goerbig}). In addition, a multi-fractal behavior has been observed for the wavefunctions at the transition from a localized to delocalized regime in disordered electronic systems~\cite{Morgenstern, Richardella, Evers}. However, these systems do not allow one to study the influence of non-integer dimensions on the electronic properties. Geometric electronic fractals, in which electrons are confined to a self-similar fractal geometry with a dimension between one and two, have only been studied from a theoretical perspective. For these fractals, a recurrent pattern in the density of states as well as extended and localized electronic states were predicted~\cite{Domany, Ghez, Andrade1989, Wang}. Recently, simulations of quantum transport in fractals revealed that the conductance fluctuations are related to the fractal dimension~\cite{Katsnelson}, and that the conductance in a \Spi fractal shows scale-invariant properties~\cite{Chakrabarti, Liu, Lin}.

Here, we report how to construct and characterize, in a controlled fashion, a fractal lattice with electrons: the electrons that reside on a Cu(111) surface are confined to a self-similar \Spi geometry through atomic manipulation of CO molecules on the Cu(111) surface. The manipulation of surface-state electrons by adsorbates has been pioneered by Crommie \emph{et al.}\cite{Crommie} and has been used to create electronic lattices "on demand", such as molecular graphene~\cite{Manoharan}, an electronic Lieb lattice~\cite{Drost2017,Slot2017}, a checkerboard and stripe-shaped lattice~\cite{Otte}, and a quasiperiodic Penrose tiling~\cite{Collins2017}. We characterized the first three generations of an electronic \Spi triangle by scanning tunneling microscopy and spectroscopy, acquiring the spatially and energy-resolved electronic local density of states (LDOS). These results were corroborated by muffin-tin calculations as well as tight-binding simulations based on artificial atomic $s$-orbitals coupled in the \Spi geometry.

\begin{figure}[ht!]
\includegraphics[width=0.70\textwidth]{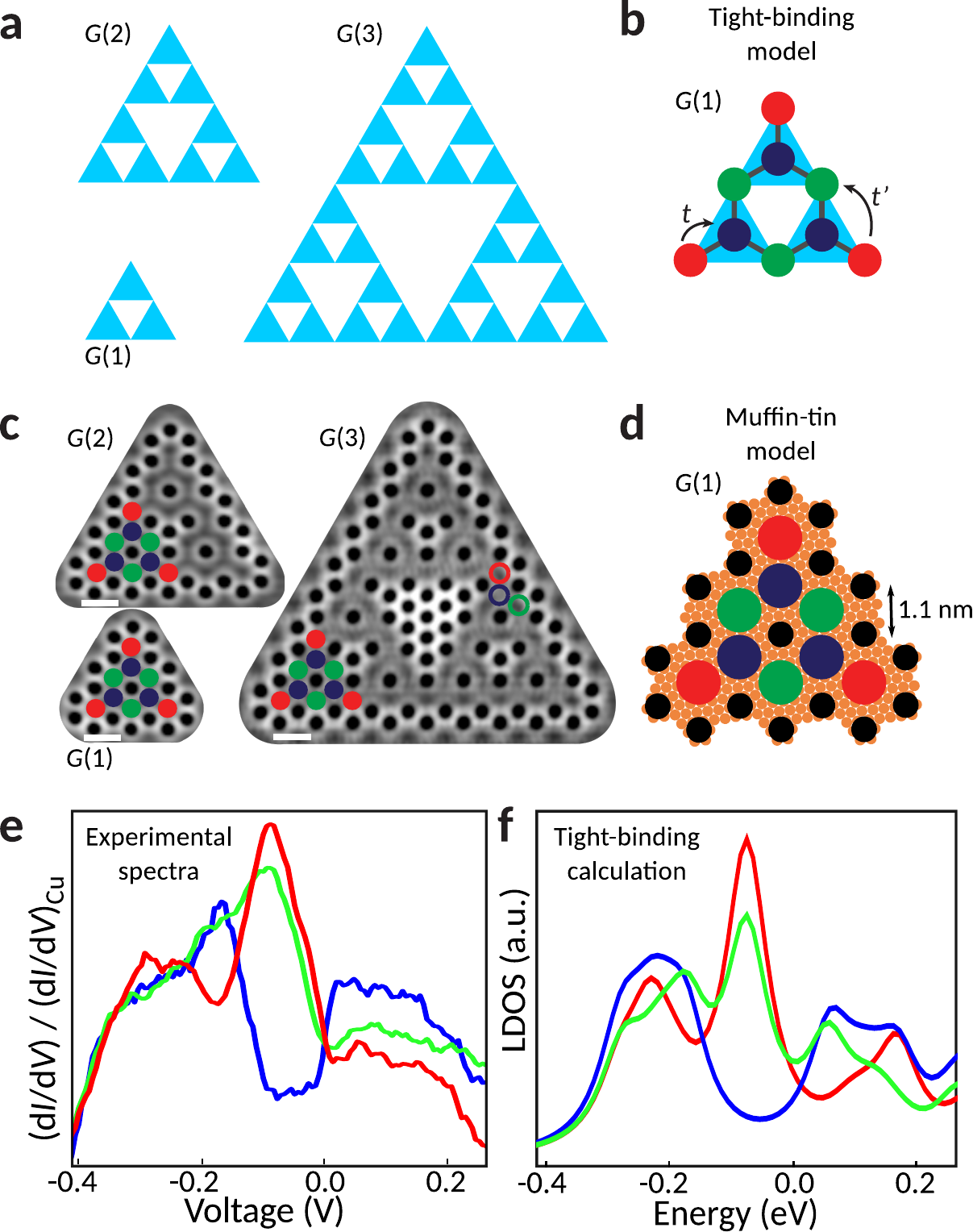} \caption{\textbf{Geometry of the \Spi triangle fractal}.\scriptsize \textbf{a}, Schematic of \Spi triangles of the first three generations $G(1)$-$G(3)$. $G(1)$ is an equilateral triangle subdivided in four identical triangles, from which the center triangle is removed. Three $G(1)$ ($G(2)$) triangles are combined to form a $G(2)$ ($G(3)$) triangle. \textbf{b}, Geometry of a $G(1)$ \Spi triangle with red, green and blue atomic sites. $t$ and $t'$ indicate nearest-neighbor and next-nearest-neighbor hopping between the sites in the tight-binding model. \textbf{c}, Constant-current STM images of the realized $G(1)$-$G(3)$ \Spi triangles. The atomic sites of one $G(1)$ building block are indicated as a guide to the eye. Imaging parameters: $I = 1\,$nA, $V = 1\,$V for $G(1)- G(2)$ and $0.30\,$V for $G(3)$. Scale bar, $2$ nm. \textbf{d}, The configuration of CO molecules (black) on Cu(111) to confine the surface-state electrons to the atomic sites of the \Spi triangle. \textbf{e}, Normalized differential conductance spectra acquired above the positions of red, blue and green open circles in \textbf{c} (and equivalent positions). \textbf{f}, LDOS at the same positions, simulated using a tight-binding model with $t=0.12\,$eV, $t'=0.01\,$eV and an overlap $s=0.2$.\label{configuration}}
\end{figure}

\normalsize The \Spi triangle with Hausdorff dimension $\log(3)/\log(2)=1.58$  is presented in Fig.~\ref{configuration}a. We define atomic sites at the corners and in the center of the light blue triangles as shown in Fig.~\ref{configuration}b for the first generation $G(1)$~\cite{Shang,Oftadeh}: $G(1)$ has three inequivalent atomic sites, indicated in red, green and blue, which differ by their connectivity. A triangle of generation $G(N)$ consists of three triangles $G(N-1)$, sharing the red corner sites. The surface-state electrons of Cu(111) are confined to the atomic sites by adsorbed CO molecules, acting as repulsive scatterers. Fig.~\ref{configuration}c shows the experimental realization of the first three generations of the \Spi triangle and Fig.~\ref{configuration}d shows the relation with the artificial atomic sites. The distance between neighboring sites is $1.1\,$nm, such that the electronic structure of the fractal will emerge in an experimentally suitable energy range~\cite{Manoharan}.

Figure~\ref{configuration}e presents the experimental LDOS at the red, blue and green atomic sites in the $G(3)$ \Spi triangle (indicated by the open circles in Fig.~\ref{configuration}c). The differential conductance (d$I$/d$V$) spectra were normalized by the average spectrum taken on the bare Cu(111) surface, similar to Ref.~[\!\!\citenum{Manoharan}]. The onset of the surface-state band is located at $V = -0.45\,$V. We focus on the bias window between $-0.4\,$V and $0.3\,$V. Around $V = -0.3\,$V the LDOS on the red, green and blue sites is nearly equal, whereas slightly above $V= -0.2\,$V, the red sites exhibit a distinct minimum, while the green and blue sites show a considerably higher LDOS. At $V = -0.1\,$V, the blue sites show a minimum, while the red and green sites exhibit a pronounced maximum in the LDOS. At $V = +0.1\,$V, the blue sites show a larger peak in the differential conductance, while the green and red sites exhibit a smaller peak. The experimental LDOS is in good agreement with both the tight-binding (see Fig.~\ref{configuration}f) and muffin-tin simulations (see Supplementary Information). This finding corroborates that our design leads to the desired confinement of the 2D electron gas to the atomic sites of the \Spi geometry. In addition, it allows us to characterize the wavefunctions of the chosen \Spi geometry in detail.

\begin{figure} \centering
\includegraphics[width=0.85\textwidth]{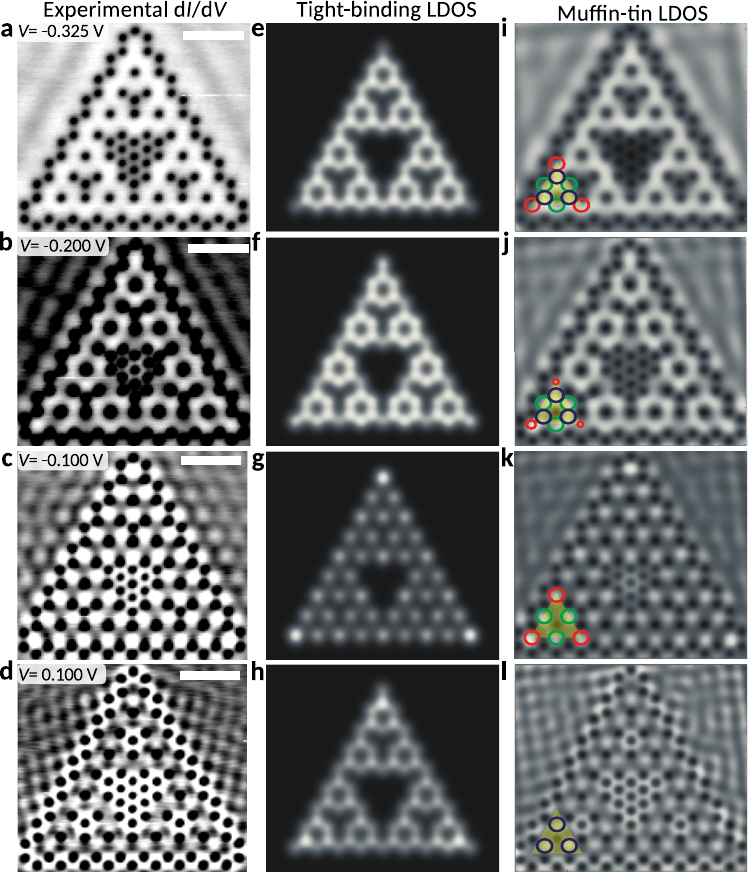}
\caption{\textbf{Wavefunction mapping}. \textbf{a-d}, Differential conductance maps acquired above a $G(3)$ \Spi triangle at bias voltages $-0.325\,$V, $-0.200\,$V, $-0.100\,$V, and $+0.100\,$V. Scale bar: $5\,$nm. \textbf{e-h}, LDOS maps at these energies calculated using the tight-binding model. \textbf{i-l}, LDOS maps simulated using the muffin-tin approximation. As a guide to the eye, a $G(1)$
building block is indicated, in which a larger radius of the circles corresponds to a larger LDOS at an atomic site, while no circle indicates a node in the LDOS. \label{maps}}
\end{figure} 
Figure 2 shows experimental wavefunction maps obtained at different bias voltages and a comparison with simulations using a tight-binding and muffin-tin model. In a thought experiment, we will discuss how electrons can be transported across the setup between a source and a drain at arbitrary positions. At a bias voltage of $-0.325 \,$V, the red (R), green (G) and blue (B) sites all have a high LDOS, and this also holds between the sites. Hence, from a chemical perspective, this wavefunction has strong bonding character, yielding an excellent conductivity from source to drain along (R-B-G-B-R)-pathways. At $V = -0.2\,$V, the red sites that connect the $G(1)$ triangles have a low amplitude: the wavefunction of the $G(3)$ triangle partitions into nine parts, each corresponding to a $G(1)$ triangle. The self-similar \Spi geometry thus leads to a subdivision of a fully bonding wavefunction delocalized over the $G(3)$ \Spi triangle at $-0.325 \,$V in self-similar $G(1)$ parts at $-0.2\,$V, demonstrating self-similar properties of the LDOS itself. At the latter bias voltage, the conductivity along (R-B-G-B-R)-pathways suffers from the lower amplitude on the red sites (except the red corner sites). At $V = -0.1\,$V, the LDOS shows a marked minimum on the blue sites and a peak at the green and red sites. From the tight-binding calculation, we find that the wavefunction has nodes on the blue sites, corresponding to a non-bonding molecular orbital from a chemical perspective. It is clear that the conductivity along the (R-B-G-B-R)-pathway mediated by nearest-neighbor hopping has vanished, and that electrons have to perform next-nearest-neighbor hopping between the red and green sites to propagate. These results connect with the theoretically calculated transmission of a \Spi carpet on a hexagonal lattice, which exhibits a gap in the conductivity although there is a high DOS in the system~\cite{Katsnelson}. Finally, at $V = +0.1\,$V, all blue sites in the $G(3)$ \Spi structure have a high amplitude, whereas the red and green sites exhibit a low amplitude. Again, the conductivity between source and drain is suppressed. We note that the LDOS maps of the three generations $G(1)-G(3)$ show the same features (see Supplementary Information), which is a consequence of the self-similarity of the geometry. We study this scale-invariance of the wavefunction in more detail with the box-counting method.

\begin{figure}[h!]
\centering
\includegraphics[width=1.0\columnwidth]{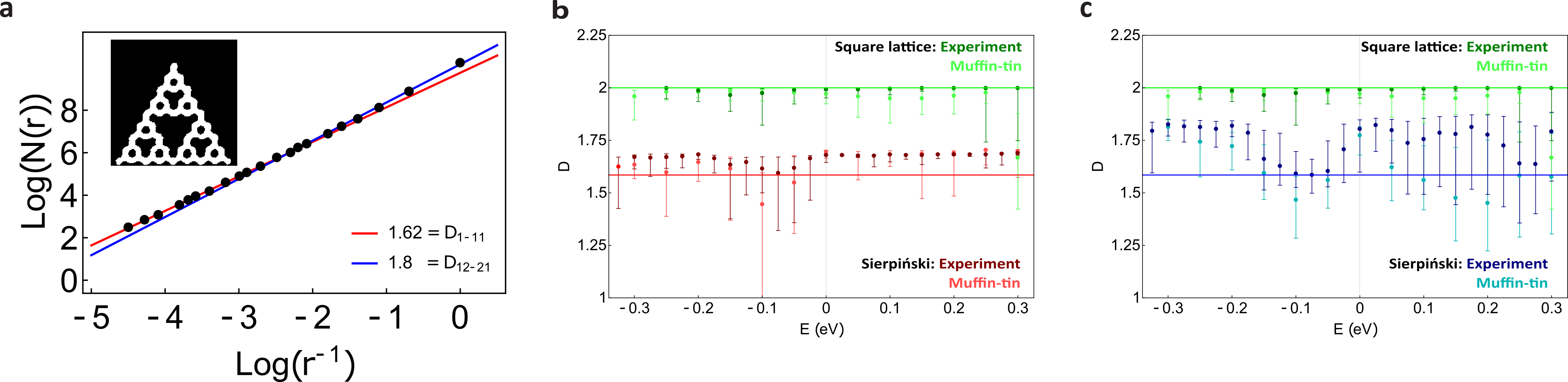}
\caption{\textbf{Fractal dimension of the \Spi wavefunction maps.} \textbf{a}, The box-counting dimension of the experimental wavefunction map acquired at $V = -0.325 \,$V is obtained from the slope of $\log(N)$ vs. $\log(r^{-1})$ where either the first 11 datapoints starting from the left (largest boxes, red slope) or the last 10 datapoints (smallest boxes, blue slope) are taken into account. Inset: binarized and masked map of the experimental wavefunction obtained with a binarization threshold of 65\%. \textbf{b}, Determination of the fractal dimensions of the LDOS of the $G(3)$ \Spi triangle (red) using the 11 largest boxes (red slopes) and comparison with the 2D square lattice from Ref.~[\!\!\citenum{Slot2017}] (green) for the experimental (dark) and muffin-tin (light) wavefunction maps with a threshold of 65\%. The error bars display the error due to the choice of the binarization threshold, indicating the fractal dimension at LDOS thresholds of 55\% (top) and 75\% (bottom). The solid lines show the geometric \Spi Hausdorff dimension ($D = 1.58$) and that of the square lattice ($D=2$). \textbf{c}, The box-counting dimension of the wavefunction maps when the last 10 datapoints are taken into account for the \Spi triangle (blue).}
\label{Fig1b}
\end{figure}
In order to determine whether the electronic wavefunctions inside the \Spi structure inherit the scaling properties of the \Spi geometry, we determine the fractal dimension of the wavefunction maps at different energies following the procedure in Ref.~[\!\!\citenum{Foroutan}]. We calculate the box-counting dimension (also called Minkowski-Bouligand dimension) for both the experimental and simulated muffin-tin LDOS maps using $D= \lim_{r \rightarrow 0} \frac{\log{N(r)}}{\log({1/r})}$, with $N$ the number of squares on a square grid required to cover the contributing LDOS and $r$ the side length of these squares. In this procedure, the wavefunction maps are positioned on a 360x360 pixel square grid. The parts of the wavefunction maps that are in the regions containing an agglomeration of CO's are excluded by applying a mask. Further, we choose a threshold (55\%, 65\% and 75\% of the maximum LDOS) above which the LDOS contributes to define a binary image. Finally, the box-counting dimension of this image is calculated. The number of boxes $N$ is counted for 21 box sizes ranging from 1 to 90 pixels. Subsequently, the fractal dimension is given by the slope of the log-logplot for $N(r)$. Details can be found in the Supplementary Information. 
A typical log-logplot is presented in Fig. 3a, where the inset shows the binarized LDOS image of the experimental wavefunction map at $V = -0.325 \,$V. The red slope is calculated for the largest 11 boxes and the blue slope for the smallest 10 boxes. Fig. 3b shows the box-counting dimension obtained from these red slopes for the experimental (dark red) and theoretical (light red) wavefunction maps acquired at various energies. For comparison, we also show the dimension obtained for the wavefunction maps of a square lattice (dark and light green, for the experiment and theory respectively), realized in the same way and measured in the same energy window~\cite{Slot2017}. The difference between the experimental and simulated maps is ascribed to a more gradual contrast in the simulation, where also contributions of the tip density of states do not play a role. It can be clearly seen that the box-counting dimension of the \Spi triangle is close to the theoretical Hausdorff dimension 1.58 (red solid line), while the square lattice has a dimension close to 2 (green solid line). When calculating the box-counting dimension for the blue slopes for the \Spi triangle, we find the results shown in Fig. 3c. When using the 10 smallest boxes, the analysis yields higher values of the dimensions for the fractal, but they are still well below 2 and well below the values obtained for the square lattice. This behavior is well understood (see Supplementary Information).
From these results and an additional scaling analysis shown in the SI, we conclude that the wavefunctions inherit the fractal dimension and therefore the scaling properties of the geometry to which they are confined, and that this dimension can be non-integer.

\begin{figure}
\includegraphics[width=0.95\textwidth]{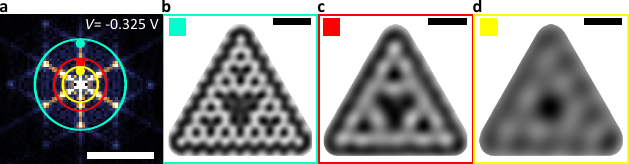}
\caption{\textbf{Fourier analysis of wavefunction maps.} \textbf{a}, Fourier transform of the experimental differential conductance map at $-0.325\,$V. The $k$-values outside the circles are excluded from the Fourier-filtered images in \textbf{b}-\textbf{d}. Scale bar: $k = 3\,$nm$^{-1}$. \textbf{b-d}, Wavefunction map at $-0.325\,$V after Fourier-filtering, including merely the $k$-values within the turquoise (\textbf{b}), red (\textbf{c}), and yellow (\textbf{d}) circles indicated in \textbf{a}. Scale bar: $5\,$nm.}
\label{fourier}
\end{figure}
Finally, we show how the self-similarity of the wavefunction maps is reflected in momentum space. The Fourier-transformed wavefunction map at $V = -0.325\,$V  (Fig.~\ref{fourier}a) exhibits distinct maxima at $k = 1.9\,$nm$^{-1}$ (turquoise), $k = 1.0\,$nm$^{-1}$ (red), and $k = 0.5\,$nm$^{-1}$ (yellow). These maxima correspond to the next-nearest-neighbor distances between the artificial atomic sites (see Fig.~\ref{configuration}), the side of a $G(1)$ triangle, and the side of a $G(2)$ triangle in real space, respectively. We then transform parts of the Fourier map back into real space (Fig.~\ref{fourier}b-d). The data inside the turquoise circle recover the full $G(3)$ \Spi triangle, as shown in Fig.~\ref{fourier}b. Transforming the values inside the red circle, however, results in a \Spi triangle of generation 2, while the size is retained (see Fig.~\ref{fourier}c). Analogously, transforming the data inside the yellow circle yields a first-generation \Spi triangle (Fig.~\ref{fourier}d). This shows that the $G(3)$ wavefunction contains Fourier terms of the prior generations. The self-similar features of the \Spi triangle are thus inherently encoded in momentum space.

We have demonstrated a rational concept of building electronic wavefunctions with a fractional dimension from artificial atomic sites that couple in a controlled way. We discussed the wavefunctions that form by coupling the $s$-orbitals of artificial atoms in the single-electron regime. While this study represents the simplest thinkable case, it already displays several aspects of fractal confinement. The emergent fractionalisation of the wavefunction at the single-particle level has profound implications and opens a series of interesting questions for future investigation: Do electrons in $D$ = 1.58 behave like Luttinger liquids? Do they exhibit the fractional quantum Hall effect in the presence of a strong perpendicular magnetic field, or is the behaviour hybrid between 1D and 2D? How does charge fractionalisation manifest when the wavefunction is itself already fractional? Recent theoretical work already addresses parts of these questions and corroborates the potential of electrons in fractal lattices, showing that the \Spi carpet and gasket host topologically protected states in the presence of a perpendicular magnetic field~\cite{Neupert}. Furthermore, the design of artificial-atom quantum dots coupled in a fractal geometry can also be implemented in semiconductor technology, thus making it possible to perform spectroscopy and transport experiments under controlled electron density. This would form a versatile platform to explore fractal electronics with several internal degrees of freedom, such as orbital type, Coulomb and spin-orbit interactions, as well as external electric and magnetic fields.

\textbf{METHODS}\\
Methods are available in the online version of the article.

\textbf{Data availability}\\
All data is available from the corresponding authors on reasonable request. The experimental and simulated images used in the box-counting analysis, as well as the code, have been published (DOI:10.24416/UU01-N90LX5) and can be downloaded from \url{https://public.yoda.uu.nl/science/UU01/N90LX5.html}.The data can be accessed using open-source tools.

\nolinenumbers
\newpage

\let\oldaddcontentsline\addcontentsline
\renewcommand{\addcontentsline}[3]{}
\bibliographystyle{naturemag} 
\bibliography{bibliography.bib}  
\let\addcontentsline\oldaddcontentsline

\textbf{Supplementary Information} is available in the online version of the article.

\textbf{Acknowledgments} We thank Guido C.P. van Miert for the discussions. We acknowledge funding from NWO via grants 16PR3245 and DDC13, as well as an ERC Advanced Grant 'FIRSTSTEP' 692691.  

\textbf{Author contributions} S.N.K. did the calculations under the supervision of C.M.S. The experiments were performed by M.R.S. with contributions from S.E.F. and S.J.M.Z. under the supervision of I.S. and D.V. All authors contributed to the interpretation of the data and to the manuscript.

\textbf{Author information} Reprints and permissions information is available at www.nature.com/reprints. Authors declare no competing interest. Correspondence and requests for materials should be addressed to C.M.S. (c.demoraissmith@uu.nl) and I.S. (i.swart@uu.nl).  

\newpage

\textbf{METHODS}\\
\textbf{Scanning tunneling microscope experiments} \\
The scanning tunneling microscopy and spectroscopy experiments were performed in a Scienta Omicron LT-STM system at a temperature of $4.5\,$K and a base pressure around $10^{-10} - 10^{-9}\,$mbar. A clean Cu(111) crystal, prepared by  multiple cycles of Ar$^{+}$ sputtering and annealing, was cooled down in the scanning tunneling microscope head. Carbon monoxide was leaked into the chamber at $p \approx 3 \cdot 10^{-8}\,$mbar for three minutes and adsorbed at the cold Cu(111) surface. A Cu-coated tungsten tip was used for both the assembly and the characterization of the fractal. The CO manipulation was performed in feedback at $I = 60\,$nA and $V = 0.050\,$V, comparable to previously reported values \cite{MeyerMethods, CelottaMethods}, and was partly automated using an in-house developed program. Scanning-tunneling-microscopy imaging was performed in constant-current mode. A standard lock-in amplifier was used to acquire differential conductance spectra ($f$ = 973$\,$Hz, modulation amplitude 0.005$\,$V r.m.s.) and maps  ($f$ = 273$\,$Hz, modulation amplitude 0.010$\,$V r.m.s.) in constant-height mode. The contrast of the experimental wavefunction maps as displayed in Fig.~2 was adjusted using the software Gwyddion. For the box-counting analysis of the experimental wavefunction maps (Fig.~3), Fourier-smoothed images were used with no further adjustments of the contrast. The Fourier analysis (Fig.~4) was performed using the same software.

\textbf{Tight-binding calculations} \\
The atomic sites in the first three generations of the \Spi triangle \cite{Sierpinski} are modeled as $s$-orbitals, for which electron hopping between nearest-neighbor and next-nearest-neighbor sites is defined. The parameters used are $e_s=-0.1$ eV for the on-site energy, $t=0.12$ eV for the nearest-neighbor hopping and $t'/t = 0.08$ for the next-nearest-neighbor hopping, similar to the values reported in Ref.~[\!\!\citenum{Manoharan}]. Furthermore, we included an overlap integral $s=0.2$ between nearest-neighbors and solved the generalized eigenvalue equation $H | \psi \rangle = E \mathcal{S} | \psi \rangle$, where $\mathcal{S}$ is the overlap-integral matrix. The LDOS is calculated at each specific atomic site and a Lorentzian energy-level broadening of $\Gamma =0.80$ eV is included to account for bulk scattering. For the simulation of the LDOS maps, the same energy-level broadening was used and the LDOS at each site was multiplied with a Gaussian wavefunction of width $\sigma = 0.65 a$, where $a = 1.1\,$nm is the distance between two neighboring sites.  

\textbf{Muffin-tin calculations}\\
The surface-state electrons of Cu(111) are considered to form a 2D electron gas confined between the CO molecules, which are modeled as filled circles with a repulsive potential of $0.9$ eV and radius $R = 0.55 a/2$. The Schr\"odinger equation is solved for this particular potential landscape, and a Lorentzian broadening of $\Gamma =0.8\,$eV is used to account for the bulk scattering.

\textbf{Box-counting method} \\
The Minkowski-Bouligand~\cite{Bouligand} or box-counting method is a useful tool to determine the fractal dimension of a certain image, but has to be handled and interpreted with care. In particular, as was shown in Ref.~[\!\!\citenum{Foroutan}], the size of the boxes needs to be chosen within certain length scales. More specifically, the largest box should not be more than 25\% of the entire image side and the smallest box is chosen to be the point at which the slope starts to deviate from the linear regime in the $\log{(N)}$ vs. $\log{(1/r)}$ plot. Due to experimental limitations, it is not always possible to fully 'block' certain areas using the CO/Cu(111) platform. Redundant features that are not part of the fractal set, such as the Friedel oscillations surrounding the \Spi triangle and the LDOS between the closely-packed COs in the centers of the $G(2)$ and $G(3)$ \Spi triangles, were removed by applying a mask (see Supplementary Information). The masks serve as a proxy for the areas that should be excluded from the future experiments using other platforms (e.g. by etching or gating those areas). Furthermore, the wavefunction maps are not binary, and therefore it is necessary to specify the LDOS threshold value above which the pixels are part of the fractal set. This binarization threshold is a certain percentage of the maximum LDOS of the masked wavefunction map at a specific energy. The error introduced by the choice of the threshold is accounted for by performing the calculation procedure for the threshold percentages of 55\%, 65\% and 75\% indicated by the top, center and bottom of the error bar, respectively (see Supplementary Information). 
In addition to the box sizes chosen in the main text, the results for other box sizes are presented in the Supplementary Information. 

\let\oldaddcontentsline\addcontentsline
\renewcommand{\addcontentsline}[3]{}

\let\addcontentsline\oldaddcontentsline

\cleardoublepage


\settocdepth{subsection}
\setcounter{equation}{0}
\setcounter{figure}{0}
\setcounter{table}{0}
\setcounter{page}{1}

\makeatletter
\renewcommand{\theequation}{S\arabic{equation}}
\renewcommand{\thefigure}{S\arabic{figure}}
\renewcommand{\bibnumfmt}[1]{[S#1]}
\renewcommand{\citenumfont}[1]{S#1}

\begin{center}
\textbf{\huge{Supplementary Information}}\\ \vspace{0.5cm}
\textbf{\Large{Design and characterization of electrons in a fractal geometry}}\\ \vspace{0.5cm}
\small{S. N. Kempkes,* M. R. Slot,* S. E. Freeney, S. J. M. Zevenhuizen, \\ D. Vanmaekelbergh, I. Swart, and C. Morais Smith}
\end{center}

\tableofcontents 
\linenumbers
\subsection{Design of the \Spi triangle}
\noindent The behavior of electrons confined to zero, one, two or three dimensions has been widely described. Previous research, in particular from the experimental point of view, has mainly been limited to these integer dimensions. Fractional dimensions, which manifest themselves in fractal structures, could open a new range of possibilities. A suitable platform to create electrons in a fractal geometry is constituted by the CO/Cu(111)-system. Here, surface-state electrons of the Cu(111) substrate are confined to a desired geometry by carbon monoxide molecules, acting as repulsive scatterers. Inspired on the pioneering work by Crommie \textit{et al.}~\citeS{CrommieCu111}, who used a similar atom manipulation approach, this method was successfully employed for the 2D honeycomb lattice~\citeS{Manoharan2}, the Lieb lattice~\citeS{Slot20172}, and quasicrystals~\citeS{Collins20172}, making it an evident candidate for the realization of flat fractal types. \\
\\
\textbf{Geometry}\\
\begin{figure}[!h]
\centering
\includegraphics[width=0.9\textwidth]{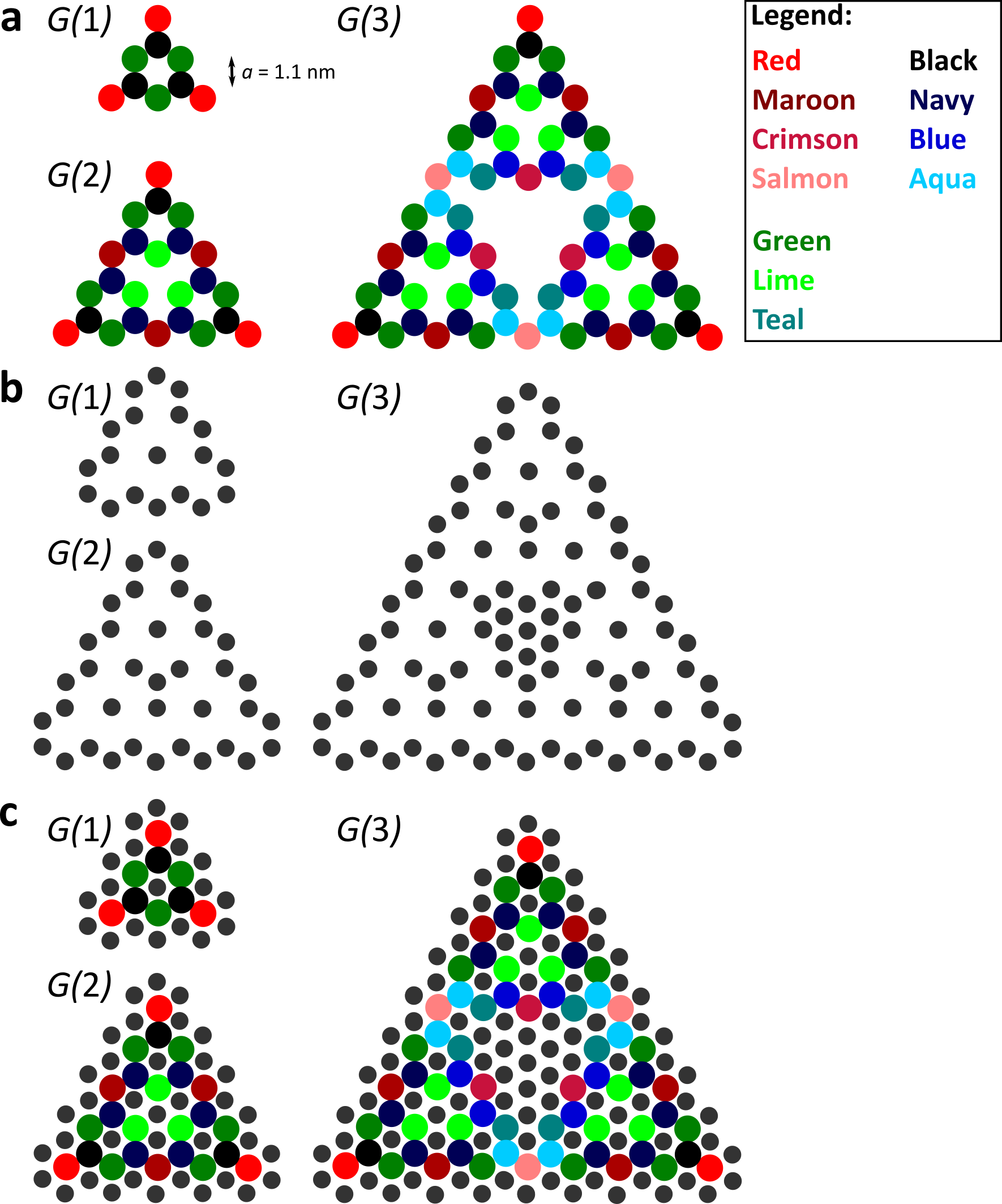}
\caption{\textbf{a}, Tight-binding geometry of the \Spi triangle with honeycomb basis for the first three generations. \textbf{b}, Configuration of CO molecules, represented by grey disks, to confine the surface-state electrons of Cu(111) to the artificial atomic sites defined in \textbf{a}. \textbf{c} The tight-binding geometry of the \Spi triangle as well as the configuration of CO molecules.}
\label{FIG:geometry}
\end{figure}
\noindent The fractal of our choice is a \Spi triangle with a honeycomb basis~\citeS{Oftadeh2, Shang2}, commensurate with the triangular symmetry of the Cu(111) surface on which the CO molecules are positioned.
Fig.~\ref{FIG:geometry}a shows the geometry of the \Spi triangle with honeycomb basis for the first three generations $G(1)-G(3)$. The $G(1)$ triangle is characterized by three inequivalent artificial atomic sites: red (connectivity 1, i.e. number of nearest-neighbor (NN) sites $z$=1), green ($z$=2), and black ($z$=3). (Note that slightly different colors are used in the Supplementary Information than in the paper to allow for a more in-depth analysis.) \\
Three $G(1)$ triangles are interconnected to form a $G(2)$ triangle. This changes the connectivity and/or environment of the initial red, green, and black sites. For instance, we distinguish between a 'red' corner site (still $z$=1) and a 'maroon' site which connects the $G(1)$ triangles (initially red site, but now with $z$=2). Similarly, there is a 'black' (still $z$=3) and a 'navy' site (still $z$=3, but with slightly different neighbors). 
Three $G(2)$ triangles, or nine $G(1)$ triangles, are interconnected to form a $G(3)$ triangle. Here, a similar distinction is made for the no longer equivalent sites. Whereas the differences are only very subtle for most sites, it is important to distinguish between the 'salmon' sites connecting the $G(2)$ triangles (position of an initial red site, but with $z$=2) and the 'red' sites with their original character at the very corners of the triangle. \\
The corresponding positions of the CO molecules, which form the anti-configuration of the fractal lattice, are shown in Fig.~\ref{FIG:geometry}b. In Fig.~\ref{FIG:geometry}c, both the \Spi lattice and the configuration of the CO molecules are shown.\\
\\
\textbf{Size}\\
\noindent The energies at which the electronic states of the fractal emerge depend on the size - \textit{i.e.} the degree of confinement - of the "artificial atom" sites to which the Cu(111) surface-state electrons are confined, as reported by Gomes \emph{et al.}~\citeS{Manoharan2}. On the one hand, the features need to appear above the onset of the Cu(111) surface state at $-0.45\,$eV. On the other hand, the contribution of the bulk states is less pronounced if the energy window is chosen below $\sim 0.5\,$eV.
For these reasons, we choose the same size of the artificial atom sites as the undoped honeycomb lattice with NN distances $a = 1.1\,$nm reported by Gomes \emph{et al.}~\citeS{Manoharan2}, which has features in the range $E \approx -0.2 \dots 0.1\,$eV.\\
\\
\begin{figure}[!ht]
\centering
\includegraphics[width=0.75\textwidth]{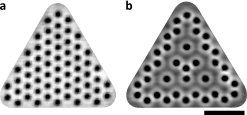}
\caption{\textbf{a}, Scanning-tunneling-microscope image of a triangle with the size of a $G(3)$ \Spi triangle, filled with a close-packed CO lattice. Imaging parameters: $I = 1\,$nA, $V=0.050\,$V. \textbf{b}, Scanning-tunneling-microscope image of the resulting $G(3)$ \Spi triangle after removing the redundant CO molecules by atomic manipulation. Imaging parameters: $I = 1\,$nA, $V=1\,$V. Scale bar: $4\,$nm.}
\label{Assembly}
\end{figure}

\noindent \textbf{Assembly}\\
\noindent In order to assemble the \Spi triangles, the CO molecules are laterally manipulated to the target copper atoms one by one (manipulation settings $I = 60\,$nA and $V = 0.050\,$V), assisted by in-house developed automated atom manipulation software. We first create an entirely filled triangle (see Fig.~\ref{Assembly}a). The periodicity of this close-packed lattice is exploited to ensure that the CO molecules are placed exactly at their desired positions, resulting in a defect-free structure. Subsequently, the redundant CO molecules are removed using the same lateral manipulation procedure (Fig.~\ref{Assembly}b). The \Spi triangles consist of 19, 37, and 94 CO molecules for the first, second, and third generation, respectively.
\newpage
\subsection{First generations: Experimental results}
\subsubsection{Energy resolution}
\label{broadening}
The resolution of the differential conductance spectra and maps is mainly limited by the used modulation amplitude of $0.005-0.010\,$V r.m.s. and the CO-induced coupling between the Cu(111) surface state and bulk states. To compare the tight-binding and muffin-tin results with the experiment, we use a linewidth $\Gamma = 0.080\,$eV, which is the same as used for the Lieb lattice geometry in Ref.~[\!\!\citenum{Slot20172}]. Note that as the number of wavefunctions scales with the number of artificial atom sites, the individual wavefunctions in \Spi triangles of higher generations become too close in energy to be resolved experimentally. This will lead to similar LDOS spectra for the \Spi triangles of generation 1-3, as shown in section D. Analogously, generations higher than 3 would have yielded similar experimental results in the used setup while requiring an increasing number of CO molecules.

\subsubsection{Differential-conductance spectra}
\begin{figure}[!h]
\centering
\includegraphics[width=0.82\textwidth]{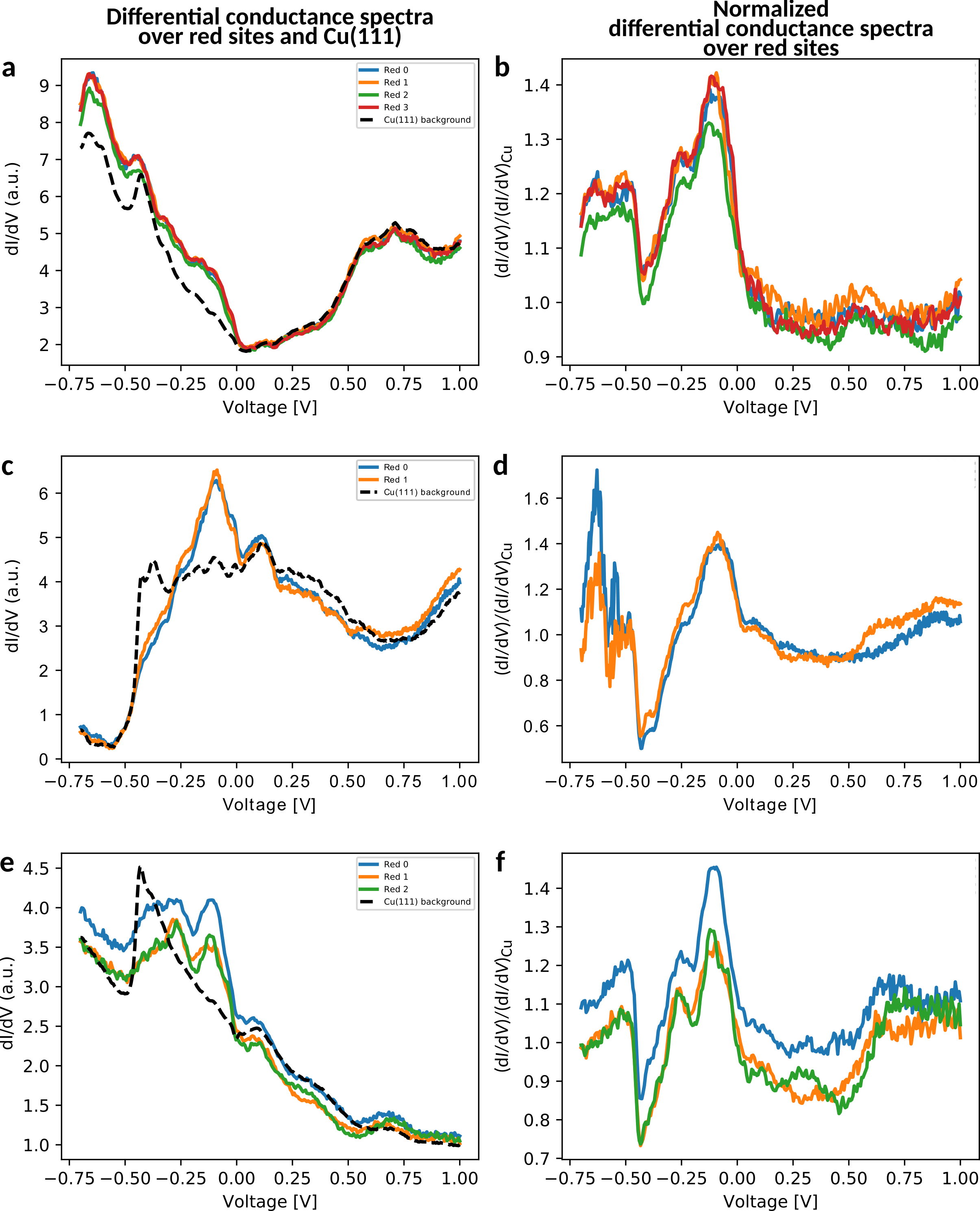}
\caption{\textbf{a}, Differential-conductance spectra acquired over several equivalent red sites in a $G(1)$ \Spi triangle (colored, solid lines) and an average of spectra over clean Cu(111) (black, dashed line). \textbf{b}, The spectra over red sites are divided by the average Cu(111) spectrum. 
\textbf{c-d}, \textbf{e-f}: same as \textbf{a-b} for tips characterized by a different spectrum on clean Cu(111). This normalization procedure leads to similar characteristic features, independently of the tip.
}
\label{SpectraNormalized}
\end{figure}
Differential-conductance spectra were acquired above the \Spi triangle and normalized by the average spectrum above clean Cu(111), following the procedure by Gomes \emph{et al.}~\citeS{Manoharan2}. This normalization cancels contributions due to the tip and the slope of the Cu(111) surface state. In Fig.~\ref{SpectraNormalized}, the normalization is shown for three different tips, characterized by significantly different averaged spectra on clean Cu(111) (black dashed lines in Fig.~\ref{SpectraNormalized}a,c,e). With each tip, we acquired several spectra over red corner sites, as shown in Fig.~\ref{SpectraNormalized}a,c,e for a $G(1)$ \Spi triangle. These spectra were divided by the Cu(111) spectrum (Fig.~\ref{SpectraNormalized}b,d,f). Similar features are observed for all different tips, corroborating the reproducibility of the normalized differential conductance spectra. \\ 
\begin{figure}[!h]
\centering
\includegraphics[width=1.00\textwidth]{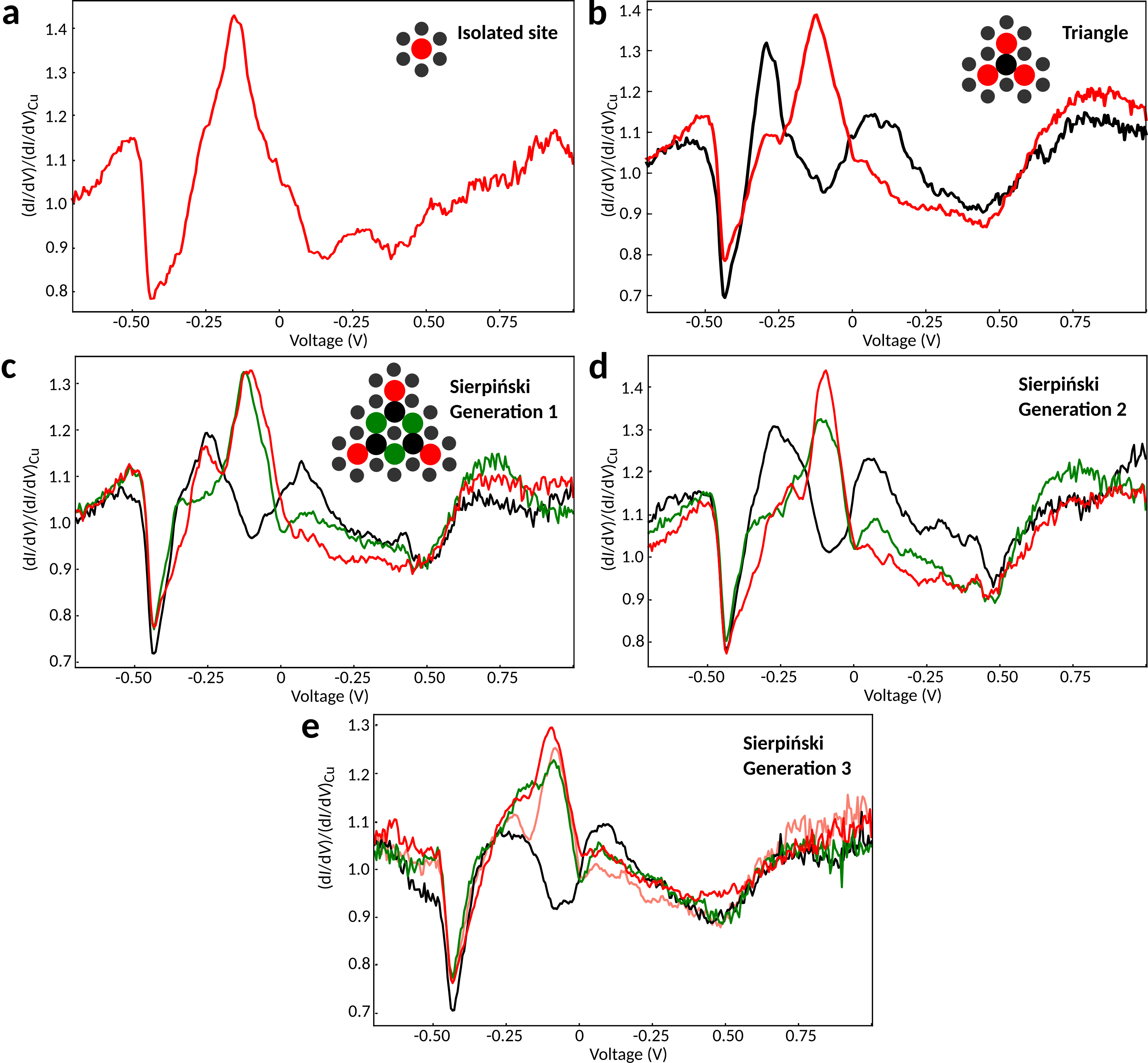}
\caption{Normalized differential-conductance spectra over an isolated artificial atom site (\textbf{a}), a simple triangle (\textbf{b}, building block of the \Spi triangle), and a $G(1)$ (\textbf{c}), $G(2)$ (\textbf{d}) and $G(3)$ \Spi triangle (\textbf{e}). The colors of the spectra correspond to the colors of the atomic sites indicated in Fig.~\ref{FIG:geometry}\textbf{a}.}
\label{spectraexperiment}
\end{figure}

In Figs.~\ref{spectraexperiment}a and \ref{spectraexperiment}b, we present a typical normalized differential-conductance spectrum on an isolated artificial atom site and on a simple triangle, which is the building block of our \Spi triangle. Furthermore, Fig.~\ref{spectraexperiment}c-e shows typical differential-conductance spectra over the red, green and black sites (defined in Fig.~\ref{FIG:geometry}a) of a $G(1)$, $G(2)$, and $G(3)$ \Spi triangle. Each \Spi triangle was built and measured at least twice with different tips. We observe that the spectrum on the isolated artificial atom site already resembles the spectra over red sites in the \Spi triangles, despite the reduced connectivity $z = 0$ (Fig.~\ref{spectraexperiment}a). The red corner sites of the simple triangle ($z$=1) also show a behavior similar to the red \Spi sites, which have the same connectivity $z$=1 (Fig.~\ref{spectraexperiment}b). Similarly, the spectrum over the black center site of the simple triangle ($z=3$) resembles the spectra over black sites in the \Spi triangles (also $z=3$). The red, green, and black \Spi sites defined in Fig.~\ref{FIG:geometry}a retain their connectivity for all generations, leading to spectra with similar features for $G(1)$-$G(3)$ (Fig.~\ref{spectraexperiment}c-e). When the connectivity of a site is changed in a higher-generation triangle, this is reflected in the spectrum, as shown for the spectrum over a 'salmon' site ($z=2$) compared to a 'red' site ($z=1$) in $G(3)$ (Fig.~\ref{spectraexperiment}e). \\ 
\begin{figure}[!h]
\centering
\includegraphics[width=0.6\textwidth]{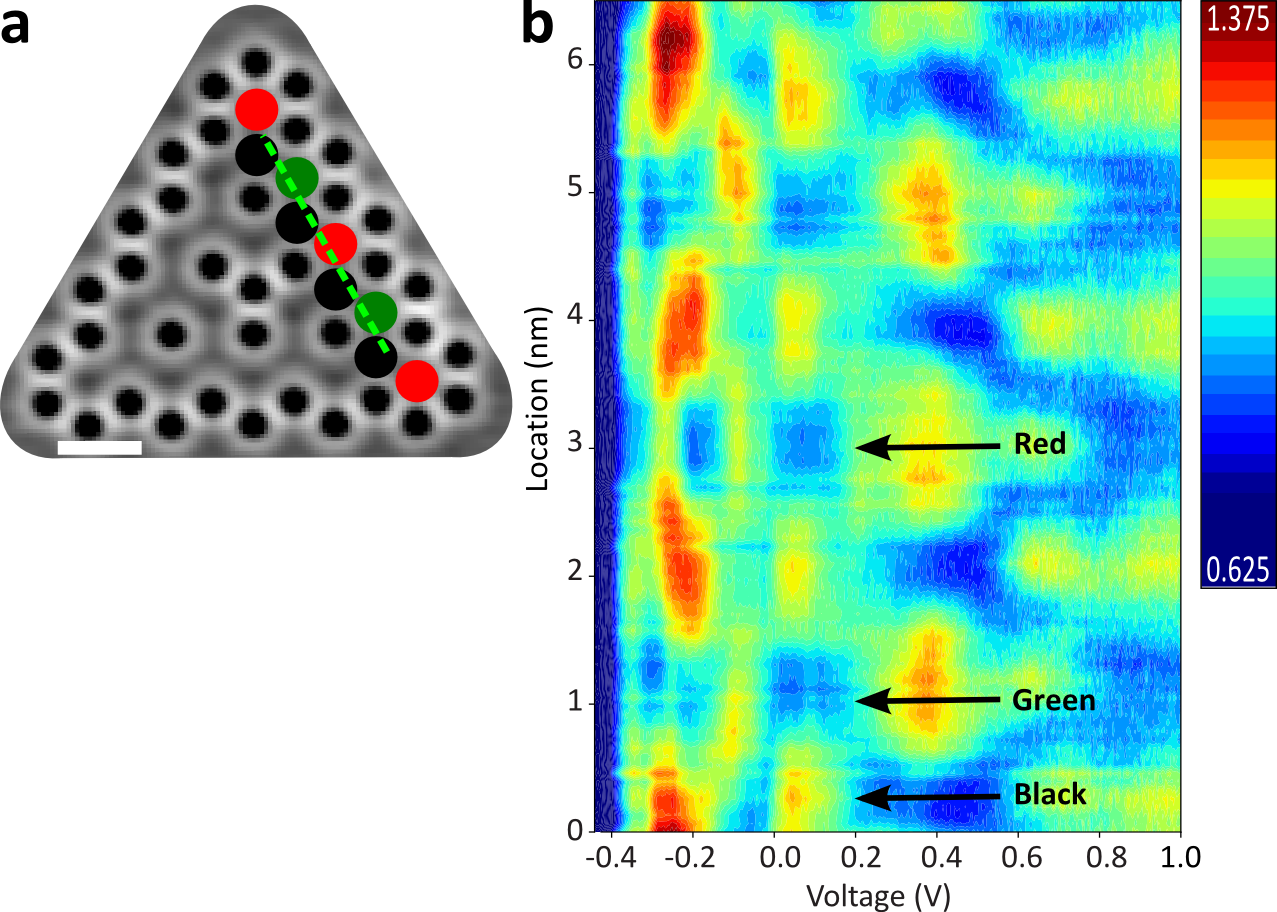}
\caption{\textbf{a}, Scanning-tunneling-microscope scan of a $G(2)$ \Spi triangle. \textbf{b}, Contour plot of 100 normalized differential-conductance spectra acquired along the green dashed line in \textbf{a}. The arrows indicate a black, green, and red site on the line.}
\label{spectraline}
\end{figure}
In Fig.~\ref{spectraline}a, a green dashed line is defined, along which a series of differential-conductance spectra was taken over the $G(2)$ \Spi triangle. The contour plot in Fig.~\ref{spectraline}b shows the amplitude of the normalized differential-conductance spectra as a function of the location along this line and the bias voltage. The main features expected for red, black and green sites are reproduced along the line. (For clarity, red and maroon as well as black and navy sites have not been distinguished in Fig.~\ref{spectraline}a.)

\subsubsection{Wavefunction maps}
\begin{figure}[!h]
\centering
\includegraphics[width=0.95\textwidth]{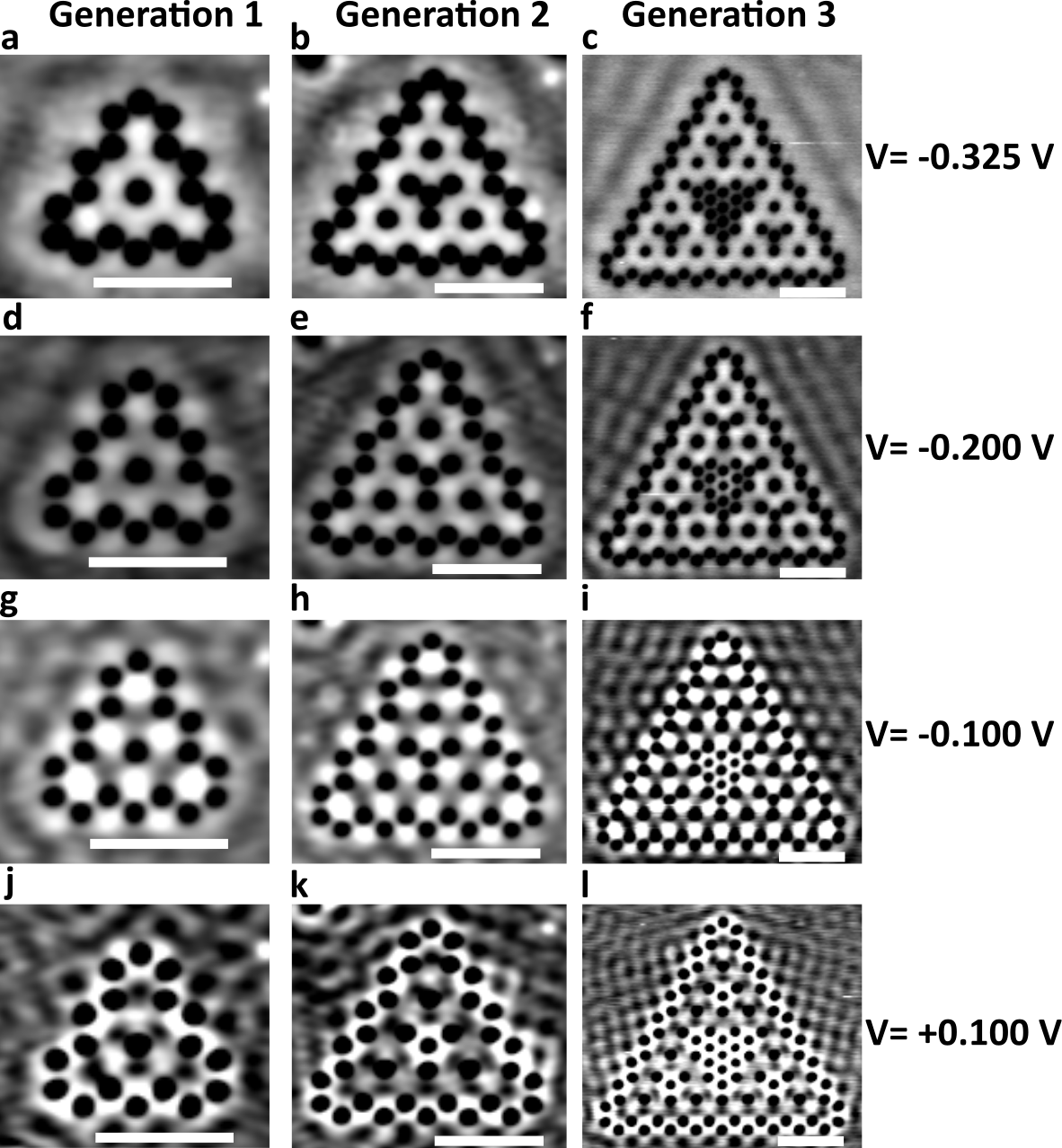}
\caption{Experimental wavefunction maps for the first three generations acquired at $V= -0.325$ V (\textbf{a-c}), $V= -0.200$ V (\textbf{d-f}), $V=-0.100$ V (\textbf{g-i}), and  $V=0.100$ V (\textbf{j-l}). Scale bar: $5\,$nm.}
\label{FIGlowenergyExp}
\end{figure}
An overview of the wavefunction maps at bias voltages $-0.325\,$V, $-0.200\,$V, $-0.100\,$V, and $+0.100\,$V above the $G(1)$, $G(2)$, and $G(3)$ \Spi triangle is shown in Fig.~\ref{FIGlowenergyExp}. At each particular bias voltage, the main features (as discussed in the main text) are similar for each generation.

\clearpage
\subsection{First generations: Muffin-tin model}
Now, we consider two different theoretical approaches to study the electronic structure of the \Spi fractal shown above. First, we concentrate on the muffin-tin approximation, which has shown to provide an accurate description of the CO/Cu(111)-system~\citeS{Manoharan2, Slot20172, Park2, Qiu2}. The surface-state electrons of Cu(111) form a 2D electron gas with an effective electron mass $m^{*} \approx 0.42\,m_e$~\citeS{Burgi2}. The band bottom is defined by the onset of the surface state at $E \approx E_{F} - 0.445\,$eV~\citeS{Kroger2}. In the muffin-tin calculations, the Schr\"odinger equation is solved for this 2D electron gas with CO molecules modeled as filled circles with a repulsive potential $V_\textnormal{\small{CO}}$ and effective radius $R=0.55 a/2$~\cite{Li2,Ropo2} ($a=1.1$nm). This leads to a Hamiltonian
\begin{equation}
H = -\frac{\hbar^2}{2 m_e^*} \nabla^2+ V_\textnormal{\small{CO}}(r),  
\end{equation}
where $V_{CO}(r) = 0.9\,$eV for $r < R$, while zero otherwise. A broadening of $\Gamma = 0.080\,$eV is included in the spectra and maps, to account for the hybridization with bulk states, as described in Sec.~\ref{broadening}.\\
In order to take into account the triangular shape geometry of the \Spi lattice, the calculations of the LDOS were carried out on a rectangular box and a triangular box, to account for the effect of Friedel oscillations in the surroundings. Due to the broadening of the LDOS, the results are almost identical for the two boundaries. We use the rectangular box for the displayed results.

\begin{figure}
\centering
\includegraphics[width=0.9\textwidth]{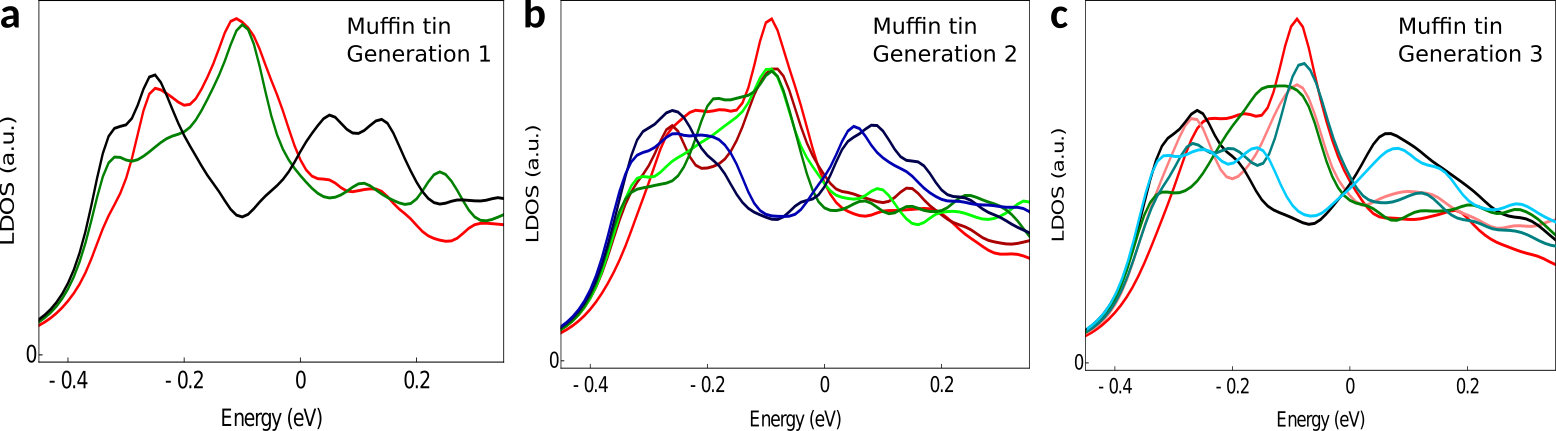}
\caption{LDOS as a function of energy obtained with the muffin-tin approach for the $G(1)$ (\textbf{a}), $G(2)$ (\textbf{b}), and $G(3)$ (\textbf{c}) \Spi triangle. The colors indicate the position of the spectra according to Fig.~\ref{FIG:geometry}. The onset of the surface-state is at $V = -0.45 \,$V. Note that the main features are similar for each generation, which is a property of the fractal structure.}
\label{spectraMT}
\end{figure}
\begin{figure}[!h]
\centering
\includegraphics[width=1.0\textwidth]{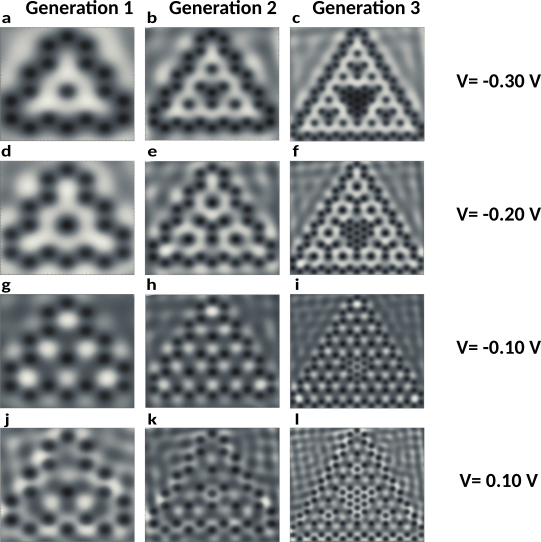}
\caption{Wavefunction maps for the first three generations of the \Spi triangle calculated within the muffin-tin approach for $V= -0.30$ V (\textbf{a-c}), $V= -0.20$ V (\textbf{d-f}), $V=-0.10$ V (\textbf{g-i}), and  $V=0.10$ V (\textbf{j-l})).}
\label{FIGlowenergyMT}
\end{figure}
Figure~\ref{spectraMT} shows the LDOS spectra simulated using muffin-tin calculations, adopting the color code defined in Fig.~\ref{FIG:geometry}a. Figure~\ref{FIGlowenergyMT} shows the wavefunction maps for the three generations and at energies corresponding to interesting peaks in the LDOS (at energies $V=-0.30$ V, $V=-0.20$ V and $V=\pm 0.10$ V). The results are in excellent agreement with the measured differential-conductance spectra and wavefunction maps shown in Fig.~\ref{FIGlowenergyExp}. 

\clearpage
\subsection{First generations: Tight-binding calculations}
\label{tightbinding}
The second theoretical approach that we use to describe the system is the tight-binding model. In the tight-binding approach, each artificial atom site is modeled by an $s$-orbital, which couples to the neighboring sites with NN hopping $t$ and next-nearest-neighbor (NNN) hopping $t'$ (see Fig.~1b). The Hamiltonian reads 
\begin{equation}
\mathcal{H} = \sum_i \epsilon_i c_i^{\dagger}c_i - t\sum_{\langle i,j\rangle}\left(c_i^{\dagger}c_j + H.c.\right) - t'\sum_{\langle\langle i,j\rangle\rangle}\left(c_i^{\dagger}c_j + H.c.\right),
\label{TBHamiltonian}
\end{equation}
where $c^{(\dagger)}_i$ are the annihilation (creation) operators for the electrons, $\epsilon_i$ is the on-site energy of the site-localized $s$ orbital, $\langle i,j \rangle$ denotes the sum over NN and $\langle \langle i,j \rangle \rangle$ the sum over NNN sites. Since the configuration has a honeycomb basis with the sizes of the undoped honeycomb lattice created by Gomes \textit{et al.}~\citeS{Manoharan2}, we consider similar values for the NN hopping amplitude $t =0.12\,$eV, a NNN hopping $t'= 0.08 t$ and an on-site energy $\epsilon_i = -0.10\,$eV. We also include an orbital overlap $s=0.2$ between NN, and therefore solve the generalized eigenvalue problem. This choice of parameters yields the best agreement with the experiment and the muffin-tin simulations.
Note that the fractal is a non-periodic structure, and therefore it is not possible to identify a band spectrum. In particular, due to scale-invariance the energy levels of $G(N)$ are related to the levels of $G(N-1)$, and it was shown that the energy spectrum shows self-similar features and has highly degenerate energy levels~\citeS{Domany2}. We will comment on this property in more detail in Sec.~S\ref{highgens}.\\
\begin{figure}[!h]
\centering
    \includegraphics[width=0.9\textwidth]{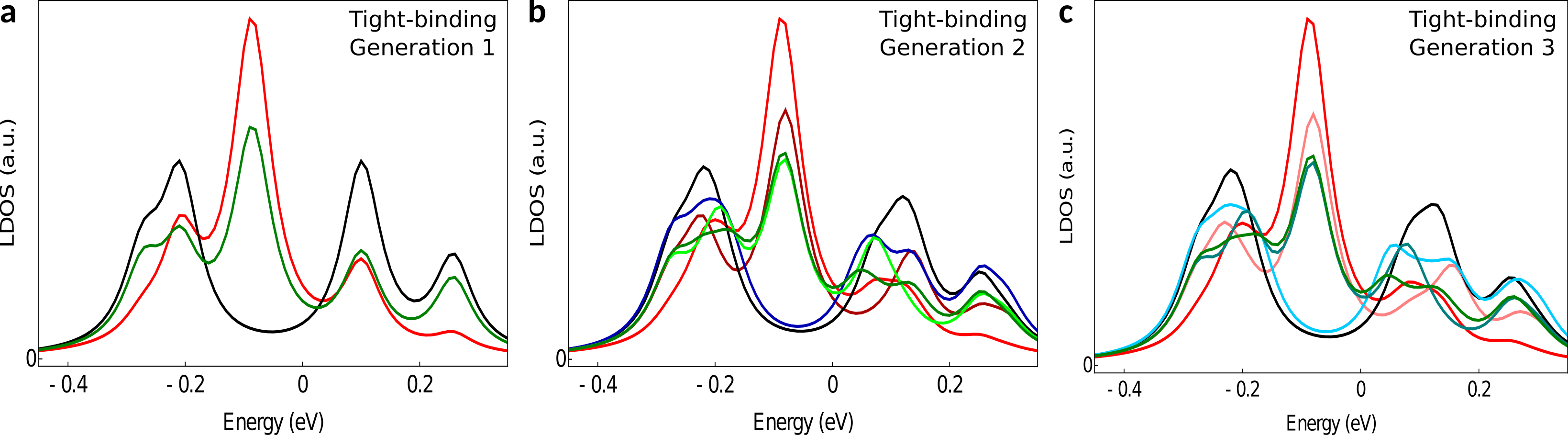}
\caption{LDOS obtained using the tight-binding approach for the $G(1)$ (\textbf{a}), $G(2)$ (\textbf{b}), and $G(3)$ (\textbf{c}) \Spi triangle, where the colors indicate the positions of the  spectra as displayed in Fig.~\ref{FIG:geometry}.}
\label{spectraTB}
\end{figure}

We solve the tight-binding eigensystem and calculate the LDOS as a function of energy via
\begin{equation} 
\text{LDOS}(\textbf{r}_0,\varepsilon)=\sum_{n} |\Phi_n (\textbf{r}_0)|^2 \delta(\varepsilon - \varepsilon_n),
\end{equation}
where the sum runs over the number of states $n$ for each lattice position \textbf{r}$_0$. In order to accommodate the broadening of the spectrum due to the repulsive scatterers, we approximate the delta function by a Lorentzian with a broadening $\Gamma= 0.080$ eV. 
In Fig.~\ref{spectraTB}, we present the LDOS as a function of energy for the $G(1)$, $G(2)$, and $G(3)$ \Spi triangle. The results are in good agreement with the experiment and muffin-tin calculations (see Figs.~\ref{spectraexperiment} and \ref{spectraMT}). In particular, we observe the features around $E=-0.3$ eV, $E=-0.2$ eV, $E=-0.1$ eV, and $E=0.1$ eV, as discussed in the main text. Moreover, the particle-hole asymmetry was accounted for by the NNN hopping, resulting in a higher intensity for the 'right' peaks around $E=0.1\,$eV than for the 'left' peaks around $E=-0.2$ eV. 
\begin{figure}[!h]
\centering
\includegraphics[width=1.0\textwidth]{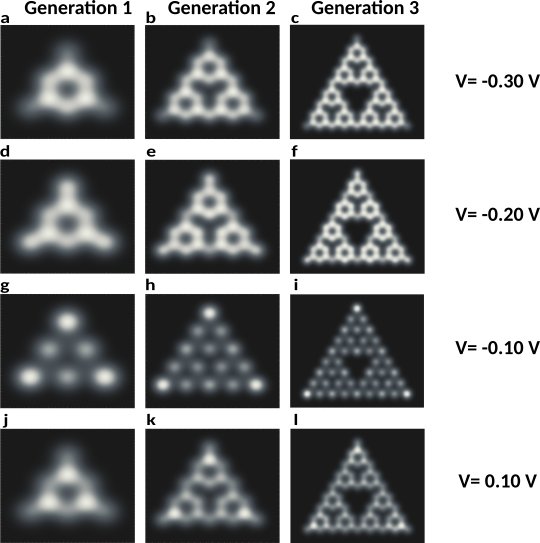}
\caption{Wavefunction maps calculated within the tight-binding model for the first three generations of the \Spi triangle at $V= -0.30$ V (\textbf{a-c}), $V= -0.20$ V (\textbf{d-f}), $V=-0.10$ V (\textbf{g-i}), and  $V=0.10$ V (\textbf{j-l}).}
\label{FIGlowenergyTB}
\end{figure}
Furthermore, we compare the wavefunction maps of the experiment and muffin-tin with the tight-binding by approximating the orbital wavefunctions as Gaussian functions. At each lattice site $i$, we model the $s$-orbitals 
\begin{equation}
\Psi(r)= \sum_i A \cdot \text{LDOS}(\textbf{r}_0, \varepsilon) \exp{\left(\frac{-(\textbf{r}-\textbf{r}_0)^2}{2\sigma^2}\right)},
\end{equation}
where the LDOS acts as an amplitude for the Gaussian wavefunction, $A$ is the normalization constant and $\sigma$ is the standard deviation, which is set by comparing with the experimental and muffin-tin wavefunction maps. The resulting figures for $\sigma= 0.65 a$ are shown in Fig.~\ref{FIGlowenergyTB} and exhibit a good correspondence with the experimental and muffin-tin maps. 
\clearpage
\subsection{Higher generations: Tight-binding calculations}
\label{highgens}
One of the interesting features of fractals is self-similarity at different length scales. In the 1980s, multiple groups used this recursiveness to derive the DOS for fractals and observed localized and extended states~\citeS{Domany2, Ghez2, Andrade19892, Andrade19912, Kappertz2, Wang2}. One specific feature is that also the DOS obeys a certain recursion relation between different generations, which is a universal feature of the fractal structure. We now investigate the self-similarity in the DOS for the \Spi fractal that is under consideration in this paper and explicitly show this repetition of the DOS using a tight-binding model.

\begin{figure}[!h]
\centering
\includegraphics[width=1.0\textwidth]{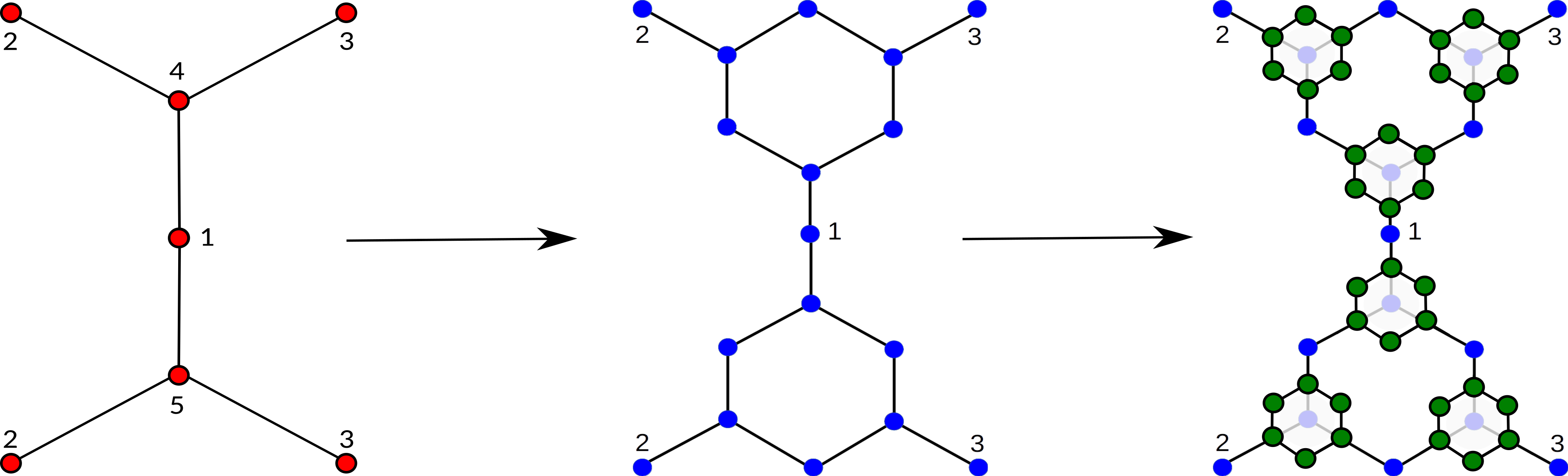}
\caption{Iteration scheme procedure. First, we double the lattice sites of our \Spi fractal to include boundary conditions. Next, we replace each site with connectivity 3 with a hexagon to go from the zeroth generation (red dots) to the first generation (blue dots) and to the second generation (green dots). This procedure is repeated each iterative step. Note that for counting the sites in this setup, we need to count all sites from the two connected \Spi triangles, and subtract 2 for the double count of sites 2 and 3. }
\label{iterationscheme2}
\end{figure}
We follow Ref.~[\!\!\citenum{Domany2}] in the discussion below. In order to construct an iteration scheme for higher-order generations, we first need to take care of the corner sites with connectivity $z = 1$, since these sites are different than the other sites in the lattice. Therefore, we mirror the image in the $x-$axis, and connect the corner sites to each other as a kind of periodic boundary conditions, see Fig.~\ref{iterationscheme2}. Now, we can distinguish two different sites: sites with connectivity $z = 2$ and sites with $z = 3$.

The corresponding Hamiltonian of this system can be separated into two subspaces $|\psi_1\rangle$ and $|\psi_2\rangle$, where the $|\psi_2\rangle$ sites are part of a hexagon, and the $|\psi_1\rangle$ connect these hexagons. The iteration scheme is then configured as follows: at each step, a hexagon replaces a site with $z = 3$, e.g. when we go from the zeroth generation (5 sites) to the first generation (15 sites), we first remove the two sites with $z = 3$ (sites 4 and 5), and insert a hexagon in that place (see Fig.~\ref{iterationscheme2}, where we go from a zeroth generation (5 sites) to a first (15 sites) and a second generation (45 sites)). Notice that the sites identified via the periodic boundary condition (two 2 sites and two 3 sites) are only counted once. A scheme is constructed in detail below while going from the first to the zeroth generation, and then reversing the process.\\
In the first generation, we want to project out the sites that are part of the two hexagons. When we only consider NN-hopping $t$, we have 
\begin{equation}
\begin{pmatrix}H_{11} & H_{12}\\
H_{21} & H_{22}
\end{pmatrix}
 \begin{pmatrix}|\psi_{1}\rangle\\
|\psi_{2}\rangle
\end{pmatrix}
 = E\begin{pmatrix}|\psi_{1}\rangle\\
|\psi_{2}\rangle
\end{pmatrix},
\end{equation}
where
\begin{equation}
H_{11}=\begin{pmatrix}0 & 0 & 0\\
0 & 0 & 0\\
0 & 0 & 0
\end{pmatrix},
  H_{12}=\begin{pmatrix}0 & 0 & 0 & -t & 0 & 0\\
0 & 0 & 0 & 0 & -t & 0\\
0 & 0 & 0 & 0 & 0 & -t
\end{pmatrix},
 H_{22}=\begin{pmatrix}0 & 0 & 0 & 0 & -t & -t\\
0 & 0 & 0 & -t & 0 & -t\\
0 & 0 & 0 & -t & -t & 0\\
0 & -t & -t & 0 & 0 & 0\\
-t & 0 & -t & 0 & 0 & 0\\
-t & -t & 0 & 0 & 0 & 0
\end{pmatrix},
 \end{equation}
and $H_{21}=H_{12}^\dagger$ for the single upward pointing triangle in generation 1 consisting of 9 sites. As a convenient choice, we assumed the on-site energy is zero. In the first step, we project out the wavefunctions $|\psi_2\rangle$ such that $H_\text{eff}| \psi_1 \rangle = \left[H_{11}+ H_{12}(E-H_{22})^{-1} H_{21}\right] | \psi_1 \rangle=E |\psi_1 \rangle$. The energy of the total lattice (consisting of 15 sites for the first generation) is bound between $\pm \sqrt{6}$. In the remainder, we focus only on the lower, upward pointing triangle (9 sites), to stay as close as possible to the decimation described in Ref.~[\!\!~\citenum{Domany2}], but a similar procedure can be done for the total lattice (which consists of an upward and downward pointing triangle as shown in Fig.~\ref{iterationscheme2}). In this upward triangle case, we obtain 
\begin{equation}
H_\text{eff}=\left(\begin{array}{ccc}
\frac{e\left(e^{2}-3\right)t}{e^{4}-5e^{2}+4} & \frac{et}{e^{4}-5e^{2}+4} & \frac{et}{e^{4}-5e^{2}+4}\\
\frac{et}{e^{4}-5e^{2}+4} & \frac{e\left(e^{2}-3\right)t}{e^{4}-5e^{2}+4} & \frac{et}{e^{4}-5e^{2}+4}\\
\frac{et}{e^{4}-5e^{2}+4} & \frac{et}{e^{4}-5e^{2}+4} & \frac{e\left(e^{2}-3\right)t}{e^{4}-5e^{2}+4}
\end{array}\right),
\end{equation}
where we introduced the dimensionless on-site energy $e=E/t$. The effective Hamiltonian describes three sites (1, 2 and 3 in Fig.~\ref{iterationscheme2}) connected with hopping $t'=-et/(e^{4}-5e^{2}+4)$ and energy $u'=2 \cdot e\left(e^{2}-3\right)t/(e^{4}-5e^{2}+4)$. The factor $2$ in $u'$ arises because each individual site (1, 2 and 3) was connected to two hexagons before the projection, and therefore the diagonal element $u'$ is twice the diagonal element of $H_\text{eff}$. To complete this step of the iteration scheme, we want to recast the on-site energy back to zero, such that we have changed nothing with respect to the original Hamiltonian (only a NN hopping) and can repeat this procedure multiple times. Therefore, we equate $e'= (E-u')/t'$, which is equivalent to $e'= - (e^2-2)(e^2-5)$. This redefined parameter now describes the energies at one lower generation.\\
\\
However, since this decimation leaves us with only three sites, whereas actually four sites should remain (one site in the center connecting the other sites), there is a final decimation step that needs to be included. This step can be written down in a similar manner as before, by decimating a 4x4 matrix into a 3x3 one as follows: we start with a 4x4 Hamiltonian, where
\begin{equation}
H_{11}=\begin{pmatrix}0 & 0 & 0\\
0 & 0 & 0\\
0 & 0 & 0
\end{pmatrix}
 , H_{12}=\begin{pmatrix}-t\\
-t\\
-t
\end{pmatrix}
 , H_{22}=\begin{pmatrix}0\end{pmatrix}
 \end{equation}
 and $H_{21}=H_{12}^\dagger$. The effective Hamiltonian is 
\begin{equation}
H_\text{eff}=\begin{pmatrix}\frac{t}{e''} & \frac{t}{e''} & \frac{t}{e''}\\
\frac{t}{e''} & \frac{t}{e''} & \frac{t}{e''}\\
\frac{t}{e''} & \frac{t}{e''} & \frac{t}{e''}
\end{pmatrix},
\end{equation}
resulting in $e'= 2- e''^2$ by following the same reasoning as before, $u''= 2 t /e''$ and $t''= -t/e''$, hence $e'= (E''-u'')/t''=2-e''^2$. Finally, we equate the two expressions for $e'$ to find $e= \pm \frac{\sqrt{7 \pm\sqrt{4 (e'')^2+1}}}{\sqrt{2}}$, which is the actual energy at a lower generation. Notice that the way in which the $\pm$ arises will make the DOS mirror symmetric around the on-site energy. Once the energy $e''$ for a low generation is known, it can be used to find the energy $e$ for a higher generation, and this starts an iterative cycle. Each energy eigenvalue in a lower generation gives rise to new eigenvalues in this way.

The iteration scheme is nearly complete. We still need to consider some special values that could not be taken into account during the iterative process: $e=0$, $e=\pm 1$ and $e=\pm 2$, because for these cases the inverse matrix is singular or the hopping $t'$ is zero. It can be shown that for these values the number of occurrences in the spectrum is $N(0)=3^n$, $N(\pm 2)= \delta_{1,n}$ and $N(\pm 1) = 5 \cdot 3^n-3^n+3^{n-1}-4 \cdot N_{n-1}$ (see Ref.~[\!\!\citenum{Domany2}]), where $n$ is the generation of \Spi triangle and $N_{n-1}$ denotes the number of eigenvalues of the previous generation. 
\begin{figure}[!h]
\begin{tikzpicture}
\node[inner sep=0pt]  at (0,0)
    {\includegraphics[width=.43\textwidth]{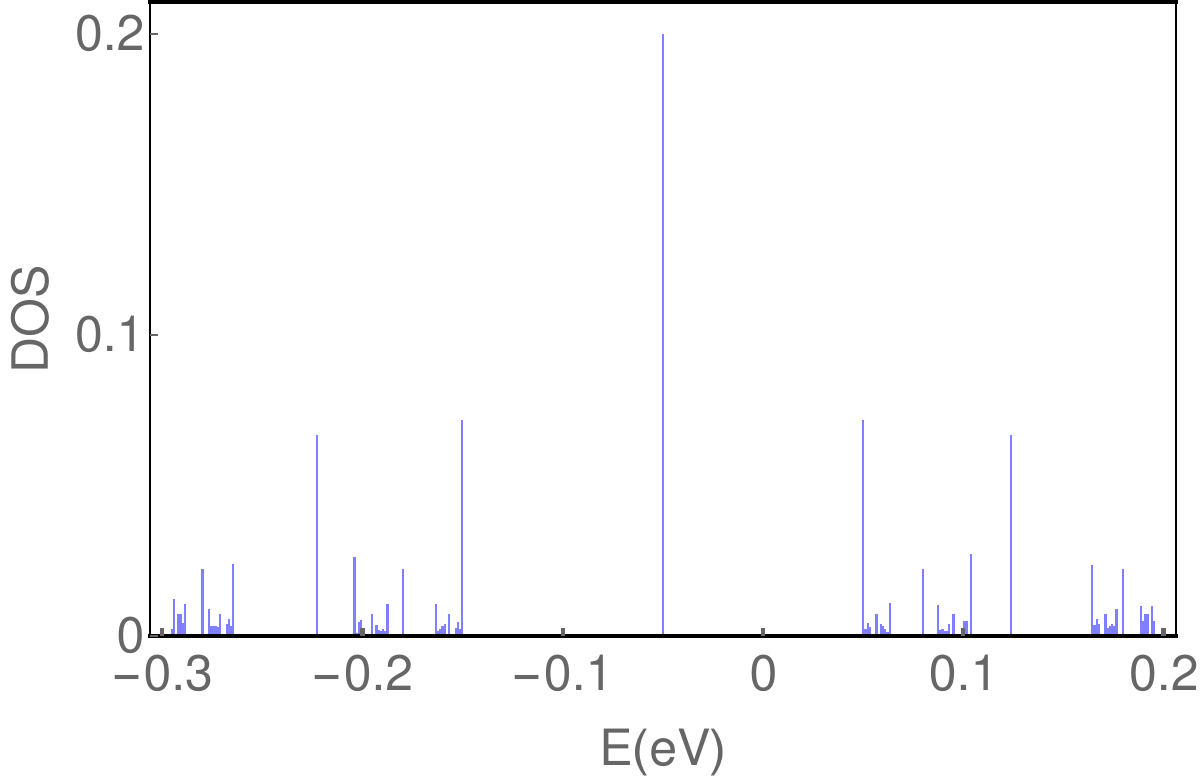}};
    \node[inner sep=0pt]  at (-3.8,2.2)
    {\textbf{a}};
\node[inner sep=0pt]  at (8,0)
    {\includegraphics[width=.43\textwidth]{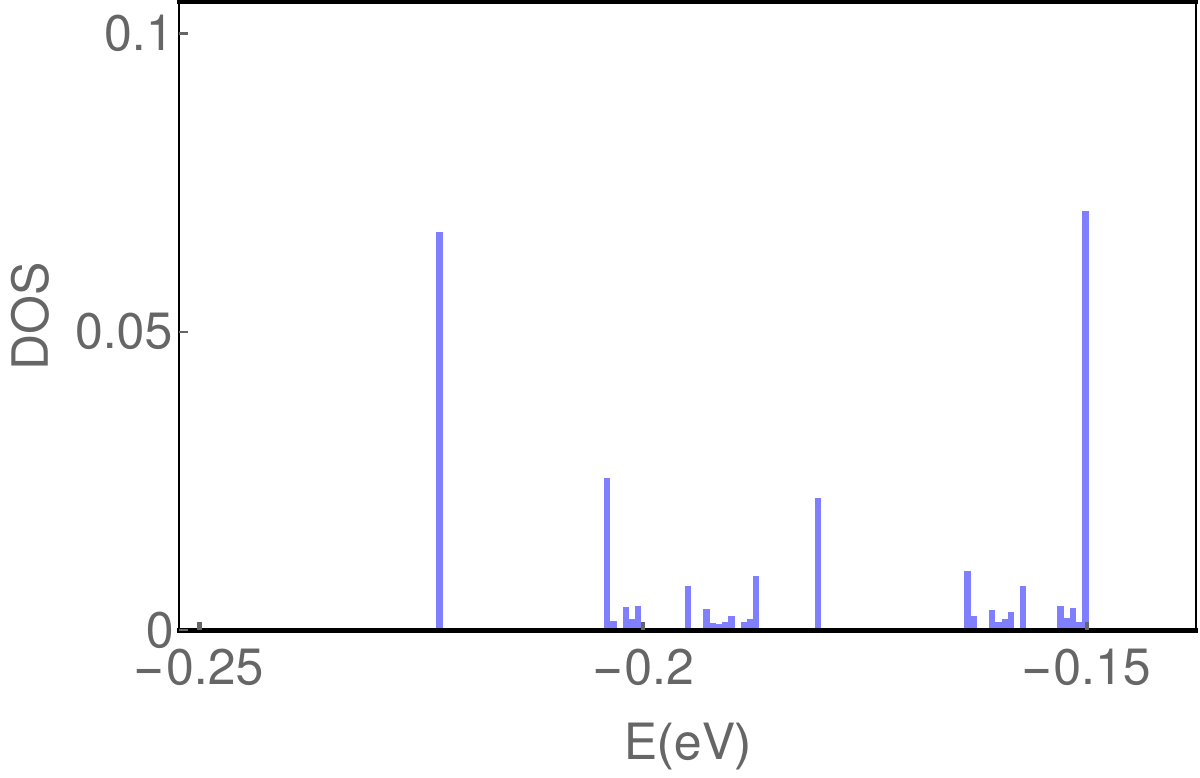}};
     \node[inner sep=0pt]  at (4.2,2.2)
    {\textbf{b}};
\node[inner sep=0pt]  at (0,-5)
    {\includegraphics[width=.43\textwidth]{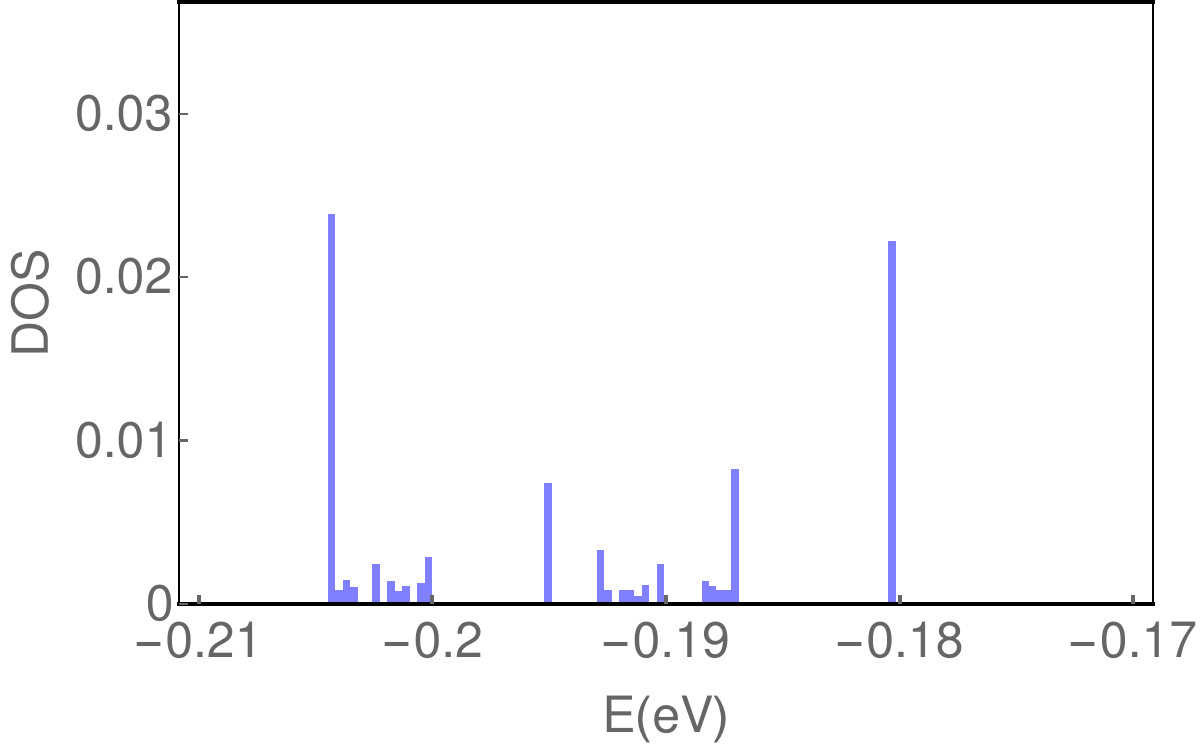}};
     \node[inner sep=0pt]  at (-3.8,-2.8)
    {\textbf{c}};
\node[inner sep=0pt]  at (8,-5)
    {\includegraphics[width=.43\textwidth]{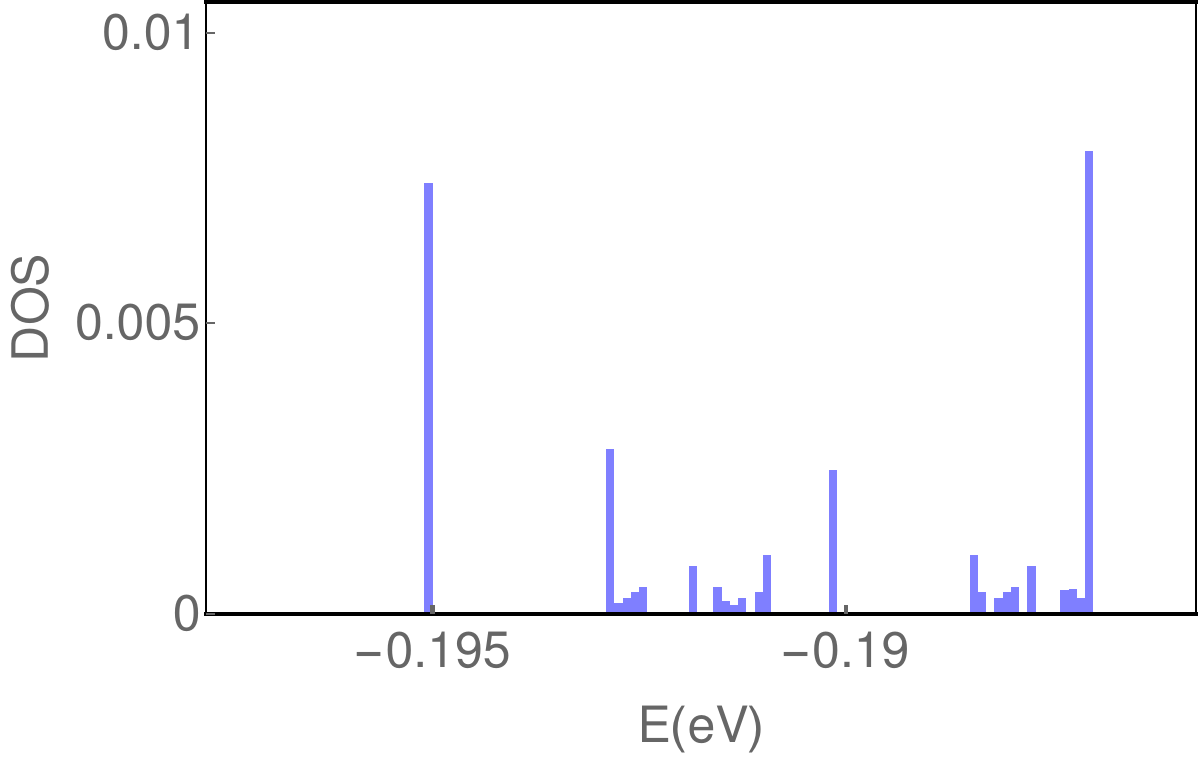}};
     \node[inner sep=0pt]  at (4.2,-2.8)
    {\textbf{d}};
\end{tikzpicture}
\caption{Fraction of DOS, where the self-similarity is clearly visible among the figures. \textbf{a}, The total fraction of DOS for generation N=10 ($5 \cdot 3^{10}$ states) and interval $\Delta E = 10^{-3}$ eV. The high central peak is due to the different connectivity of the sites and is localized at the previously mentioned green and red sites. \textbf{b, c, d}, Zoom of the DOS in the region around $E=-0.20$ eV. The self-similarity of these states is particularly clear between \textbf{b} and \textbf{d}, whereas \textbf{c} is self-similar with an additional mirror in the $y$-axis. These features are a property of the DOS of a fractal lattice. }
\label{FIGDOSiterations}
\end{figure}
The results for the DOS of the \Spi triangle are presented in Fig.~\ref{FIGDOSiterations}. Here, we changed the on-site energy for the first generation to $\epsilon =-0.05$ eV (instead of $0$ as in the discussion above) and the hopping parameter is set to $t=0.10$ eV in order to compare with the experiments, where the parameters are similar. As shown above, we calculate the eigenvalues of the 1st generation and construct the eigenvalues of the higher generations using this recursion relation. After 10 generations, we observe the DOS in Fig.~\ref{FIGDOSiterations} and see the repetitiveness in the DOS. This repeating structure is an essential feature of a fractal lattice, as was shown in Ref.~[\!\!\citenum{Domany2}].

\clearpage
\subsection{Comparison between exact solutions and experimental setup}
The exact solutions from the previous section do not take into account NNN hopping, orbital overlap and broadening, which are present in the experiment. The decimation process has not been solved for NNN hopping and overlap, as far as we are aware. However, the main features that are observed in the experiment and muffin-tin calculations are caused by the hopping parameter $t$. Due to broadening, the individual features of the repetition of the DOS cannot be resolved. Thus, we experimentally observe an LDOS that does not significantly alter with increasing generation, which is in agreement with a repetitive DOS that is subject to experimental broadening. \\ 
\begin{figure}[!h]
\centering
\includegraphics[width=1.0\textwidth]{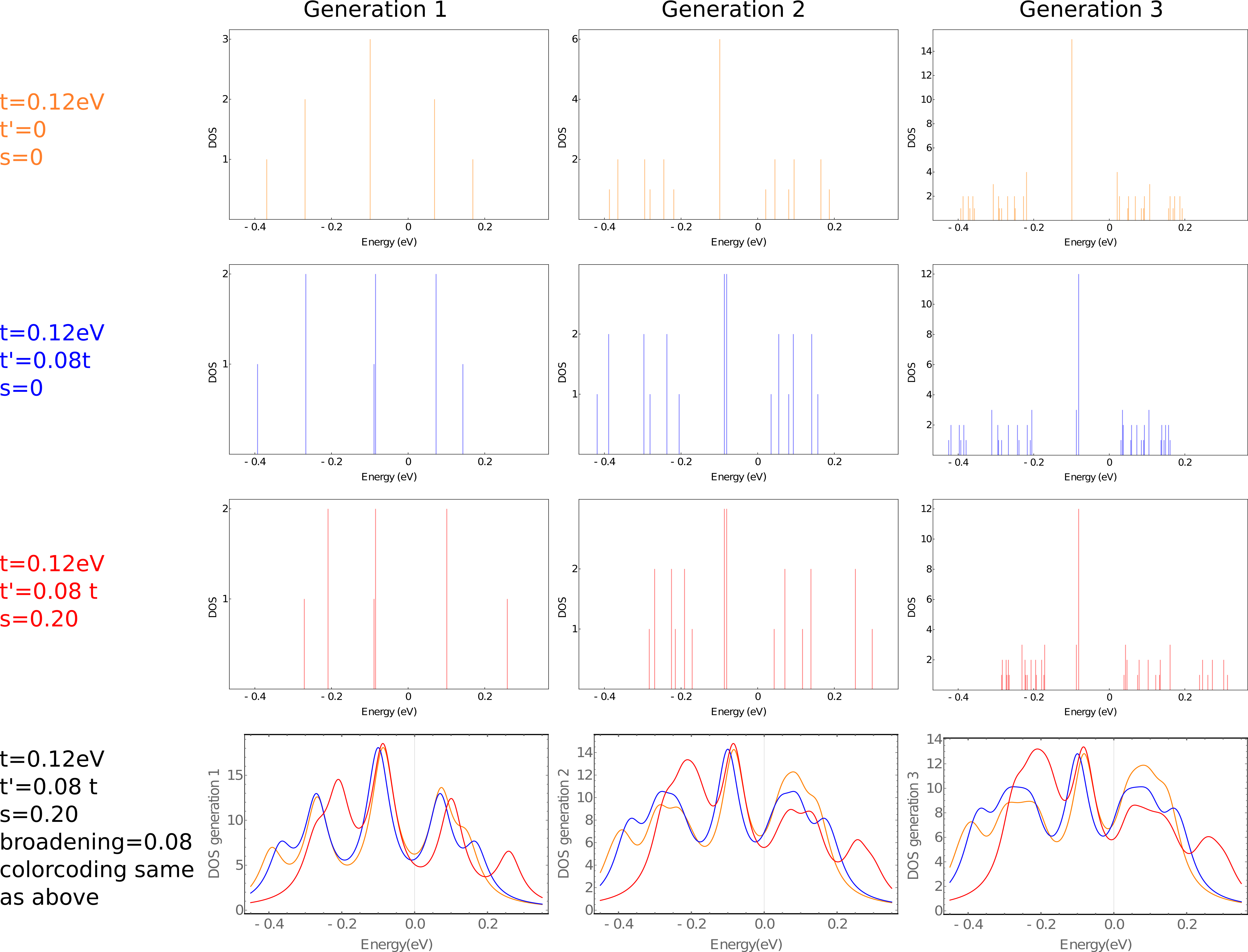}
\caption{Comparison of the DOS for the first three generations in a tight-binding model excluding NNN-hopping $t'$, orbital overlap $s$ and broadening $\Gamma$ (top line), including $t'$ (second line), including both $t'$ and $s$ (third line) and including $t'$, $s$ and broadening (last line). Without $t'$,  $s$ and broadening, we observe a symmetric DOS and we observe the self-similar structure in the third generation. When $t'$ and $s$ are included this self-similarity becomes less obvious, and when the broadening is included the features are washed away completely. The comparison between these model DOS in the last line, where the color coding is the same as in the images in the lines above, makes it clear that the significant broadening is the main parameter that prevents the self-similarity from being visible in the experiment.}
\label{tbbroadeningandoverlap}
\end{figure}
In Fig.~\ref{tbbroadeningandoverlap}, we show the calculated DOS for the first three generations with(out) NNN-hopping, overlap and broadening. It can be clearly seen from the figure that the NNN-hopping and overlap affect the peak positions and thus cloud the self-similarity. Due to the broadening, however, these detailed features can no longer be distinguished and the LDOS even shows very little differences between the generations. This similarity between the different generations is therefore an indication that the wavefunction is self-similar.

\clearpage
\subsection{Fractal dimension of the LDOS}
In the main text, a Minkowski-Bouligand dimension~\citeS{Bouligand2} - also known as box-counting or capacity dimension -  
\begin{equation}
D= \lim_{r \rightarrow 0} \frac{\log{N(r)}}{\log({1/r})}
\end{equation}
close to 1.58 was reported for the experimental and muffin-tin wavefunction maps of the $G(3)$ \Spi triangle and compared to the result of a finite 2D square lattice. In the following, we also calculate the Minkowski-Bouligand dimension for the $G(1)$ and $G(2)$ \Spi triangles. In addition, we provide an analysis addressing the various limitations of the box-counting method, in which we follow Ref.~[\!\!\citenum{Foroutan2}]. In some parts, we focus on the muffin-tin model, as this method reflects the experiment in an excellent fashion, but is not affected by experimental influences or instabilities.  
To account for reproducibility, all box-counting calculations have been performed directly on the raw data instead of using exported images. For the experimental wavefunction maps, possible glitches were suppressed using slight Fourier-smoothing in the software Gwyddion. 
To compare all images on the same level, we import the data, set the minimum to zero, mask the areas that are not part of the fractal set (including the surroundings of the triangle and the CO molecules in the center of the \Spi triangle; see the section on the influence of the use of masks below), binarize the image (using thresholds of 55\%, 65\% and 75\% of the maximum LDOS value in a masked image, chosen to represent physically meaningful extremes of which pixels should be considered as part of the wavefunction), and position the image on a square grid of 360x360 pixels. In this last part, an interpolating function between the binary points is used to position all images with comparable sizes on the same background. This means that the pixels of different structures ($e.g.$ the \Spi triangle and the square lattice) have a different size in nanometer. However, this is irrelevant for the determination of the box-counting dimension. Next, following Ref.~[\!\!\citenum{Foroutan2}], we define square boxes with pixel sizes starting with 25\% of the surrounding box length (90 pixels) down to 1 pixel. Specifically, we consider the following 21 box sizes $r$ that fit inside the $360 \,$x$360\,$ box (in number of pixels): \{90, 72, 60, 45, 40, 36, 30, 24, 20, 18, 15, 12, 10, 9, 8, 6, 5, 4, 3, 2, 1\}. Subsequently, we count the number of square boxes $N$ that contain at least one pixel that is part of the fractal set. The box-counting dimension is given by the slope in the log-logplot of $N(r)$ vs. $1/r$. In the original article, we made use of circular boxes, but now we change to the more commonly used square boxes. The counting of the boxes is based on Ref.~[\!\!\citenum{Kaurov}]. \\
\\
For a mathematical fractal, the slope of the log-logplot of $N(r)$ vs. $1/r$ is linear. However, for finite-size objects, the slope starts to deviate from linear behavior and a choice has to be made which box sizes should be taken into account in the analysis. We start by considering the largest box size and observe where the data points start to deviate from linear behavior~\cite{Foroutan2}. However, as far as we are aware, there is no objective method to decide which box sizes should be taken into account, and different choices will result in different dimensions. Too small box sizes tend towards a dimension of 2 when the image is connected and has a finite thickness ($i.e.$ it does not consist of narrow lines, but is built of parts with a finite width), since the small boxes measure also the spurious finite thickness of the lines. Therefore, the smallest box sizes are often not included in the linear regime when determining the fractal dimension. Further, sometimes the largest box size of 25\% is not a proper starting point, particularly for noisy or dispersed structures, when the linear regime sets in for smaller box sizes. To illustrate this feature, below we investigate how strongly the dimension extracted using the box-counting method depends on which box sizes are taken into account. We determine the slopes of the log-logplots, and thereby the box-counting dimension, by taking various sets of box sizes into account in the fitting procedure. We will use different regimes labeled by color to identify the dimension of the LDOS of the assembled structure. A description of the datapoints in each regime is shown in Table.~I.  

\begin{center}
\begin{table}
\begin{tabular}{ |c|c|c| } 
 \hline
 \textbf{Box sizes taken into account} & \textbf{Label} & \textbf{Color}  \\ 
 \hline
 90 - 1 pixels  & 1-21 & Green \\ 
 \hline
 90 - 15 pixels & 1-11 & Red \\ 
 \hline
 12-1 pixels & 12-21 & Blue \\
  \hline
 30-1 pixels & 7-21 & Grey\\
    \hline
 60-5 pixels & 3-17 & Orange\\
    \hline
 45-8 pixels & 4-15 & Cyan\\
 \hline
\end{tabular}
 \label{tab1}
     \caption{The different regimes for the box-counting dimension of the images}
\end{table}
\end{center}

\subsubsection{Benchmarking the box-counting analysis}

\begin{figure}[!h]
\centering
\includegraphics[width=1.0\textwidth]{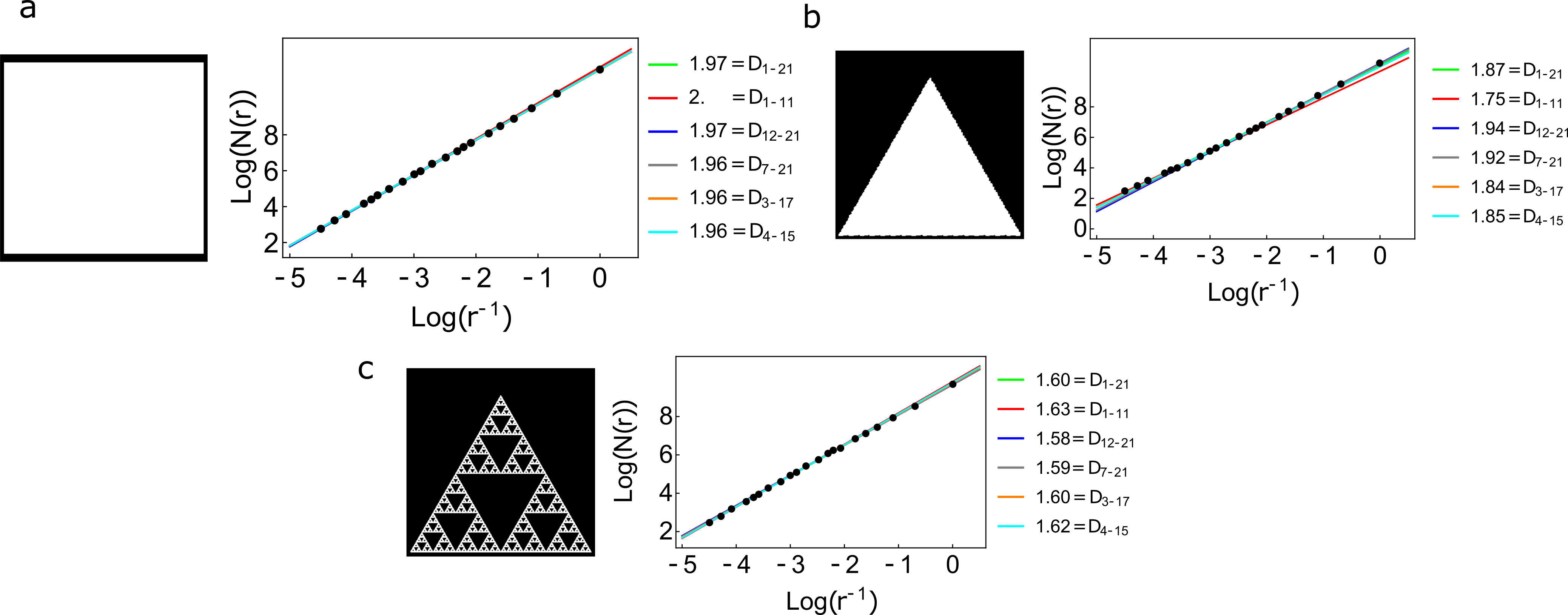}
\caption{Results of the box-counting analysis applied to a rectangle (\textbf{a}), triangle (\textbf{b}) and sixth generation \Spi triangle (\textbf{c}). The linear slope of a log-logplot of N($r$) as a function of 1/$r$ gives the fractal dimension of the white object in the image. For a rectangle, the best results are obtained when using larger boxes. In contrast, for the triangle, smaller box sizes provide better results. For a \Spi triangle, all box sizes give reasonable results.}
\label{figsquaretriangle}
\end{figure}

To benchmark the box-counting procedure, we apply it to three 2D images of systems of which the dimension is known: a rectangle and a triangle, as well as to a sixth generation mathematical \Spi triangle (created with \textit{SierpinskiMesh[10]} in Mathematica 12.0 and positioned as a binary image on the same 360x360 pixel background, resulting in an effective 6th generation \Spi triangle). All structures are positioned on the same black background of $360 \,$x$360 \,$ pixels. The images are shown in Fig.~\ref{figsquaretriangle}. The solid rectangle and triangle have the same aspect ratios as the muffin-tin calculated maps corresponding to the experiments on the square lattice and $G3$ \Spi triangle. Note that the experimental 'square' lattice of which the dimension was determined is strictly speaking a rectangular lattice (it is not possible to construct a perfect square lattice on a triangular Cu(111) surface). In this Supplementary Information, we will further refer to it as 'rectangular lattice'. Our box-counting procedure finds a dimension of $2.0$ for the solid rectangle when using the 11 largest box sizes and values close to 2.0 when considering other box sizes (see Fig.~\ref{figsquaretriangle}a). When using this range of box sizes, the dimension of the triangle is underestimated (Fig.~\ref{figsquaretriangle}b). This behavior is well known and can be understood by realizing that a square box covers more than the triangle for large box sizes. For solid 2D images, it can take some length scales for the boxes to measure the actual dimension and get the linear behavior setting in. Hence, for a solid triangle, only the small box sizes should be used to estimate the dimension (we find a value of 1.94 compared to the expected value of 2.0). For the sixth generation \Spi triangle, smaller boxes also give a (slightly) more accurate result than larger boxes ($D = 1.58$, vs. $D = 1.63$ compared to a theoretical value of $D = 1.58$). Now that we have established that the box-counting method gives reasonable results for known systems, we apply it to the experimental and theoretical LDOS maps of the artificial structures we created.

\subsubsection{Box-counting analysis of a rectangular lattice}

\begin{figure}[!h]
\centering
\includegraphics[width=1.0\textwidth]{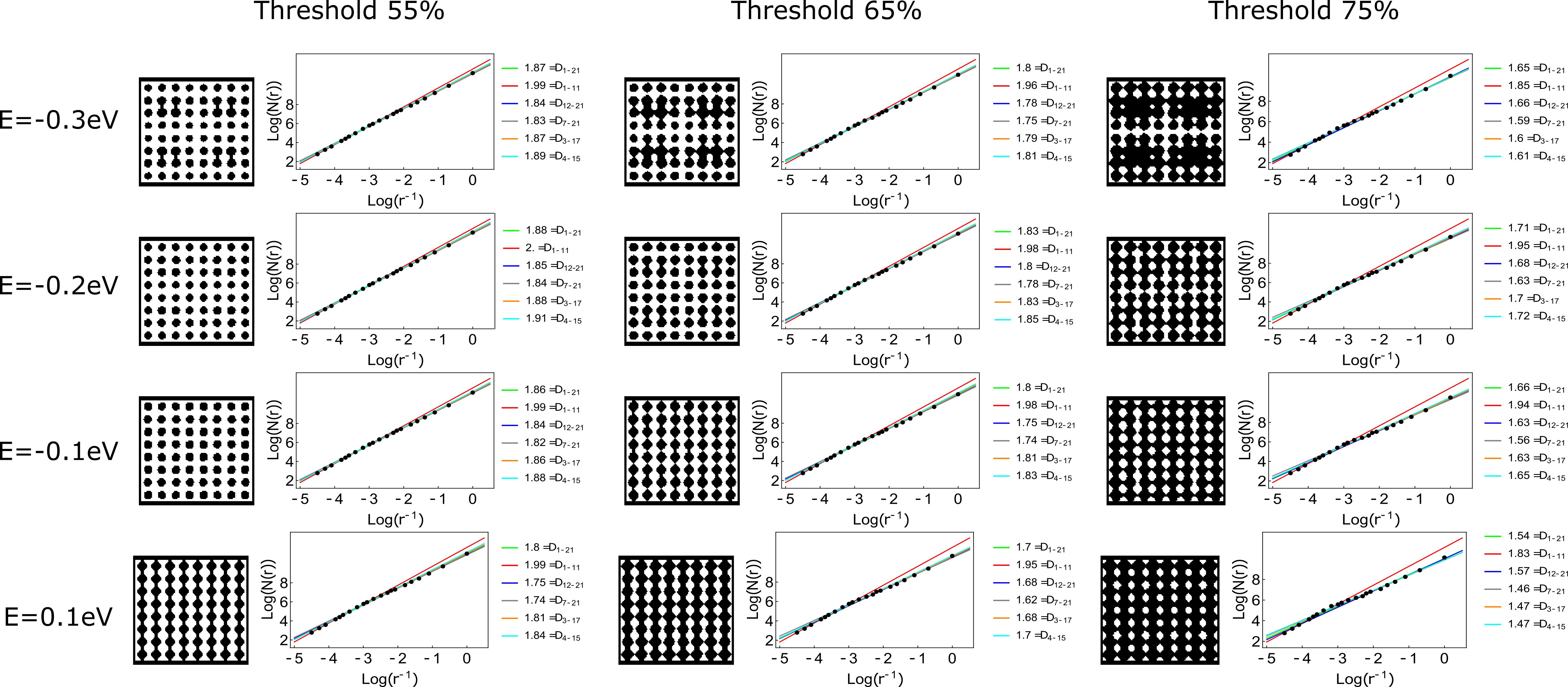}
\caption{The binary LDOS maps for the rectangular lattice obtained with the muffin-tin model for different thresholds (55\% (left), 65\% (middle) and 75\% (right)) at the energies $E=-0.30\,$eV, $E=-0.20\,$eV, $E=-0.10\,$eV and $E=+0.10\,$eV. An enclosed white area is taken into account as being part of the fractal set, which differs for a given LDOS map depending on the chosen binarization threshold. In Fig.~3 in the main text, the difference in dimension obtained for the different thresholds is indicated by the error bar. The different slopes indicate the dimensions that are found when considering the different datapoints described in Table~I.}
\label{thresholdsquare}
\end{figure}
\begin{figure}[!h]
\centering
\includegraphics[width=1.0\textwidth]{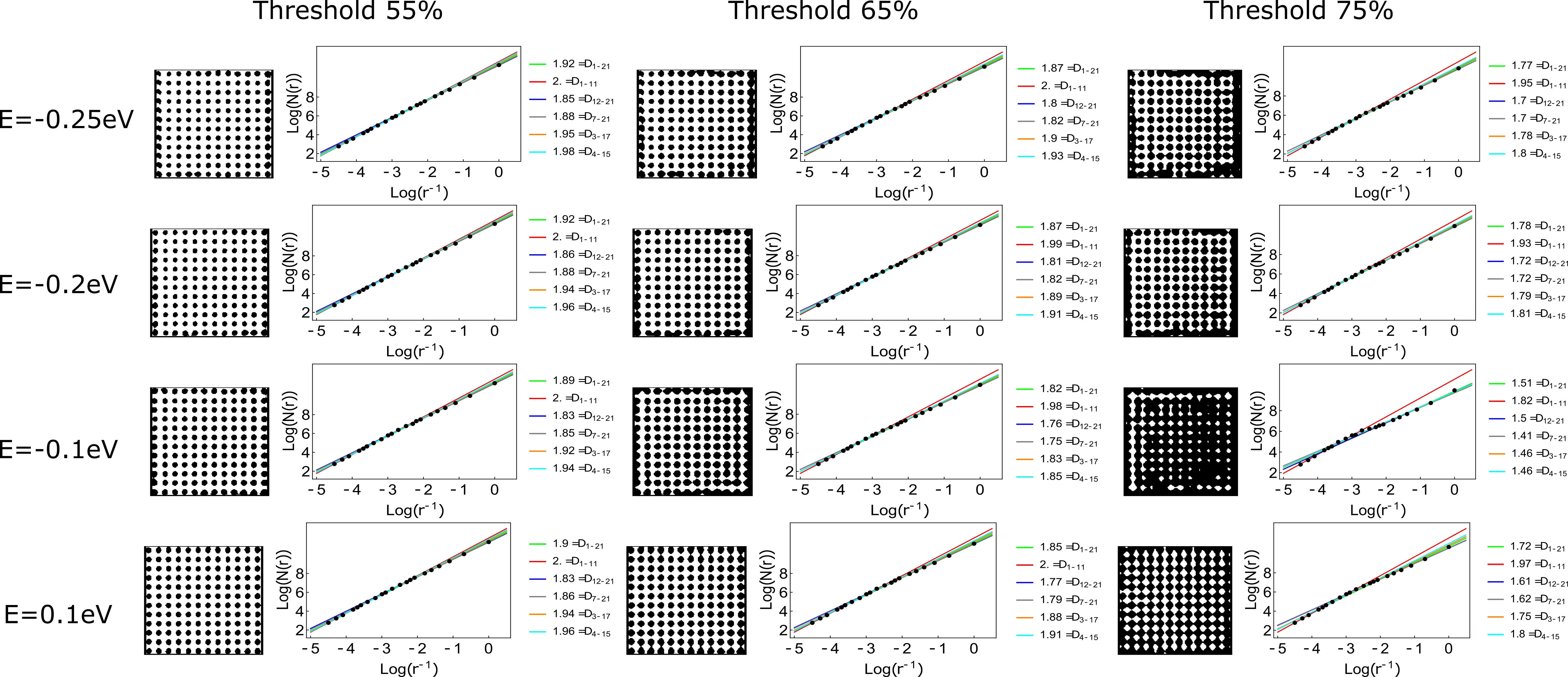}
\caption{The binary LDOS maps for the experimental rectangular lattice obtained for different thresholds (55\% (left), 65\% (middle) and 75\% (right)) at the energies $E=-0.25\,$eV, $E=-0.20\,$eV, $E=-0.10\,$eV and $E=+0.10\,$eV. An enclosed white area is taken into account as being part of the fractal set, which differs for a given LDOS map depending on the chosen binarization threshold. The difference in dimension is indicated by the error bar in Fig.~3 in the main text. The different slopes indicate the dimensions that are found when considering the different datapoints described in Table~I. }
\label{thresholdsexpsquare}
\end{figure}

We start by calculating the dimension of the rectangular lattice from Ref.~[\!\!\citenum{Slot20172}]. The results are shown in Figs.~\ref{thresholdsquare}  and \ref{thresholdsexpsquare} for the muffin-tin model and for the experimental data, respectively. Note that the small deviation from a perfect square lattice due to the underlying triangular Cu(111) geometry, as well as the finite size of the structure can lead to differences with respect to periodic square lattices. The box-counting dimension is found to depend on which box sizes are taken into account. Especially the choice between taking the 11 largest boxes (red) or the 10 smallest boxes (blue) yields different dimensions, which is most prominent for the images with a threshold of 75\%. Following the guidelines in Ref.~[\!\!\citenum{Foroutan2}], one should consider a box size starting around 25\% of the image size, and then progressively decrease the box size until the slope of the log-logplot of $N(r)$ vs. $1/r$ deviates from linear behavior. This suggests the use of the 11 largest box sizes (red lines), which roughly results in a dimension between 1.9 and 2 for this rectangular lattice. 

\subsubsection{Box-counting analysis of a \Spi triangle}

\begin{figure}[!h]
\centering
\includegraphics[width=1.0\textwidth]{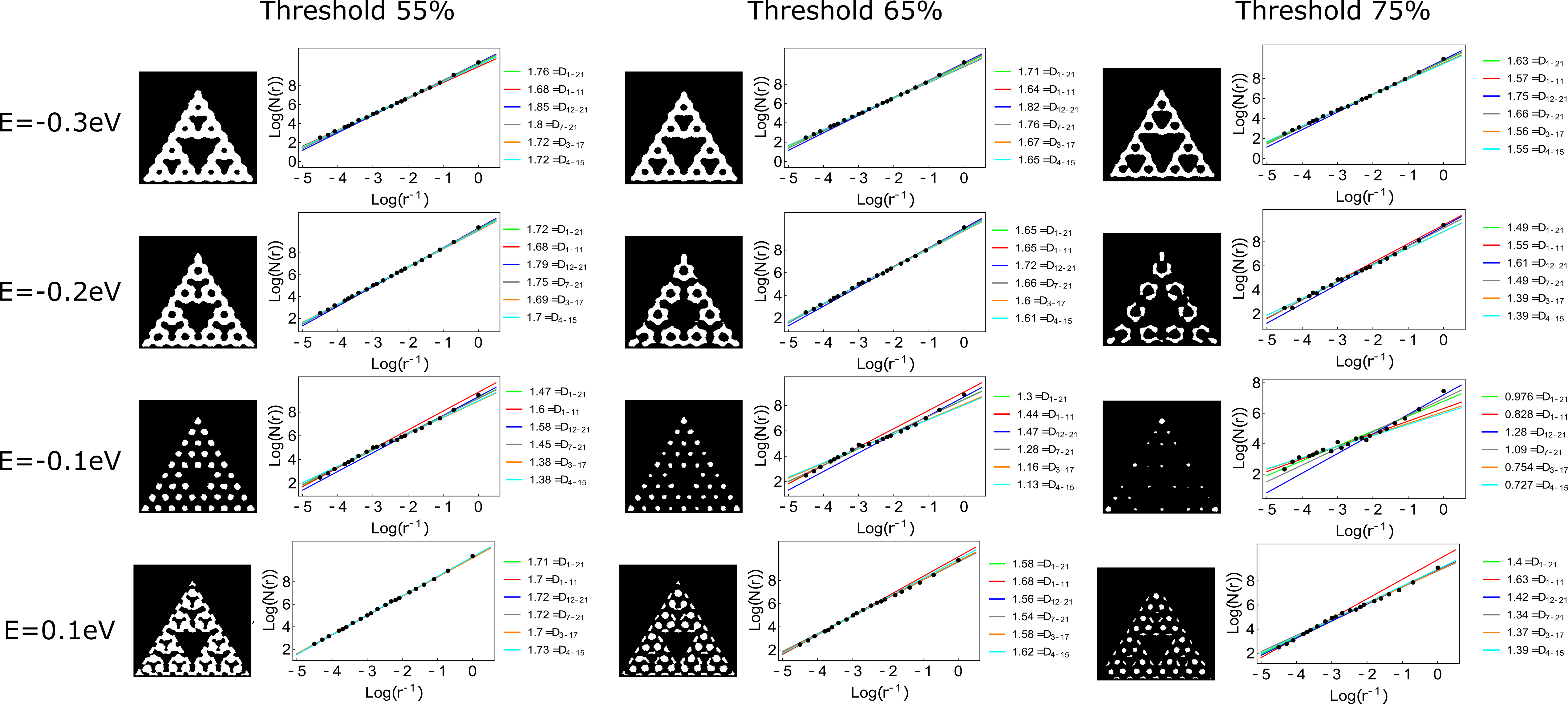}
\caption{The binary LDOS maps for the LDOS of the \Spi lattice obtained with the muffin-tin model for different thresholds (55\% (left), 65\% (middle) and 75\% (right)) at the energies $E=-0.30\,$eV, $E=-0.20\,$eV, $E=-0.10\,$eV and $E=+0.10\,$eV. An enclosed white area is taken into account as being part of the fractal set, which differs for a given LDOS map depending on the chosen binarization threshold. This difference in dimension is indicated by the error bar in Fig.~3 in the main text. The different slopes indicate the dimensions that are found when considering the different datapoints described in Table~I.}
\label{threshold}
\end{figure}
\begin{figure}[!h]
\centering
\includegraphics[width=1.0\textwidth]{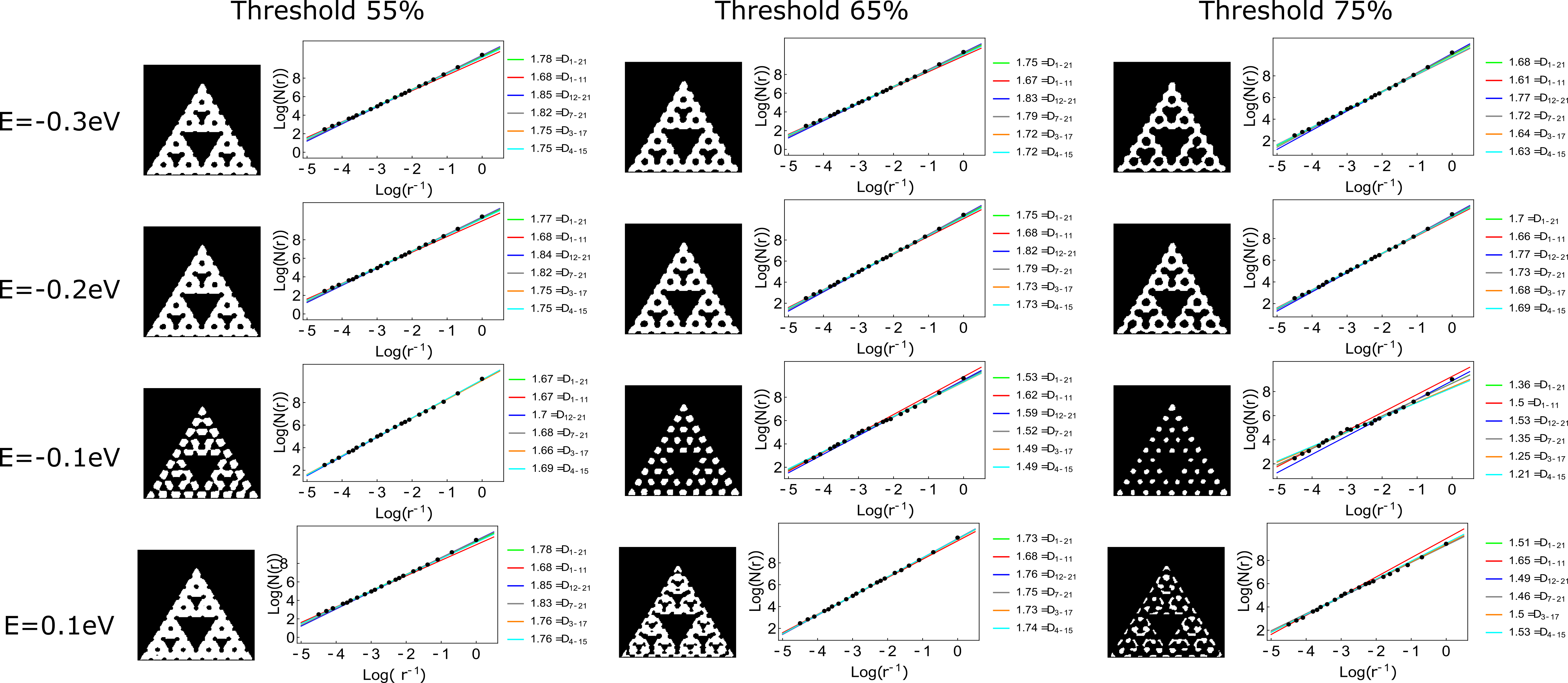}
\caption{The binary LDOS maps for the experimental \Spi wavefunction maps for different thresholds (55\% (left), 65\% (middle) and 75\% (right)) at the energies $E=-0.30\,$eV, $E=-0.20\,$eV, $E=-0.10\,$eV and $E=+0.10\,$eV. An enclosed white area is taken into account as being part of the fractal set, and this area differs depending on the chosen binarization threshold for a given LDOS map. This difference in dimension is indicated by the error bar in Fig.~3 in the main text. The different slopes indicate the dimensions that are found when considering the different datapoints described in Table~I.}
\label{thresholdsexpsierp}
\end{figure}

Figures~\ref{threshold} and \ref{thresholdsexpsierp} show the results of a similar box-counting analysis for muffin-tin and experimental LDOS maps of the $G(3)$ \Spi triangles, respectively.
Similar to the results obtained for the rectangular lattice, the dimensions are found to depend slightly on which box sizes are taken into account in the analysis. The findings on the triangle shown in Fig.~\ref{figsquaretriangle} suggest the use of the smallest box sizes in the analysis of the \Spi triangle. However, the use of small box sizes in the analysis of complex shapes can result in an overestimation of the dimension~\cite{Foroutan2}. Regardless of which box sizes are taken into account in the analysis, the dimension of the LDOS maps of the \Spi triangle is consistently below 2.0 and, importantly, below the values found for the filled triangle. The reason for this is that less smaller boxes are needed to cover the fractal set due to the presence of holes in the \Spi triangle. Hence, the box-counting method confirms that the wavefunctions have a fractal dimension.

\begin{figure}[!h]
\centering
\includegraphics[width=1.0\textwidth]{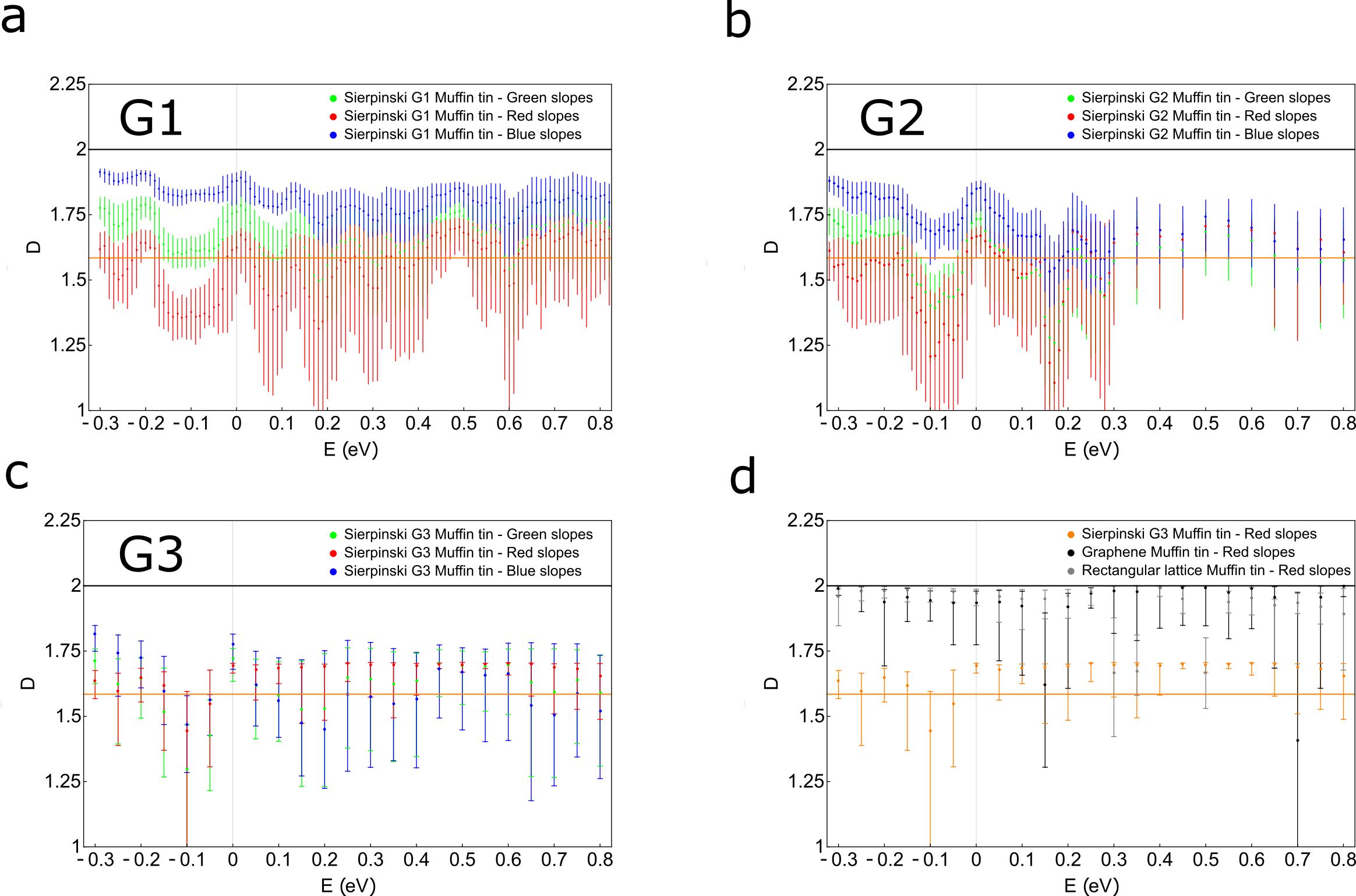}
\caption{The Minkowski-Bouligand dimension $D$ of the muffin-tin LDOS maps at different energies for $G(1)$ (\textbf{a}), $G(2)$ (\textbf{b}), and $G(3)$ (\textbf{c-d}), where in \textbf{d} we also show the results for a 2D honeycomb (black) and rectangular (grey) lattice placed on the square grid. The results in (\textbf{a-c}) are shown for the choice of different box sizes as indicated in Table.~I, while in (\textbf{d}) the red slope was chosen. The error bar displays the spread given by the choice of the threshold value for the LDOS, set to 55\%, 65\%, and 75\% for the top, center, and bottom, respectively.}
\label{fig4}
\end{figure}

\begin{figure}[!h]
\centering
\includegraphics[width=1.0\textwidth]{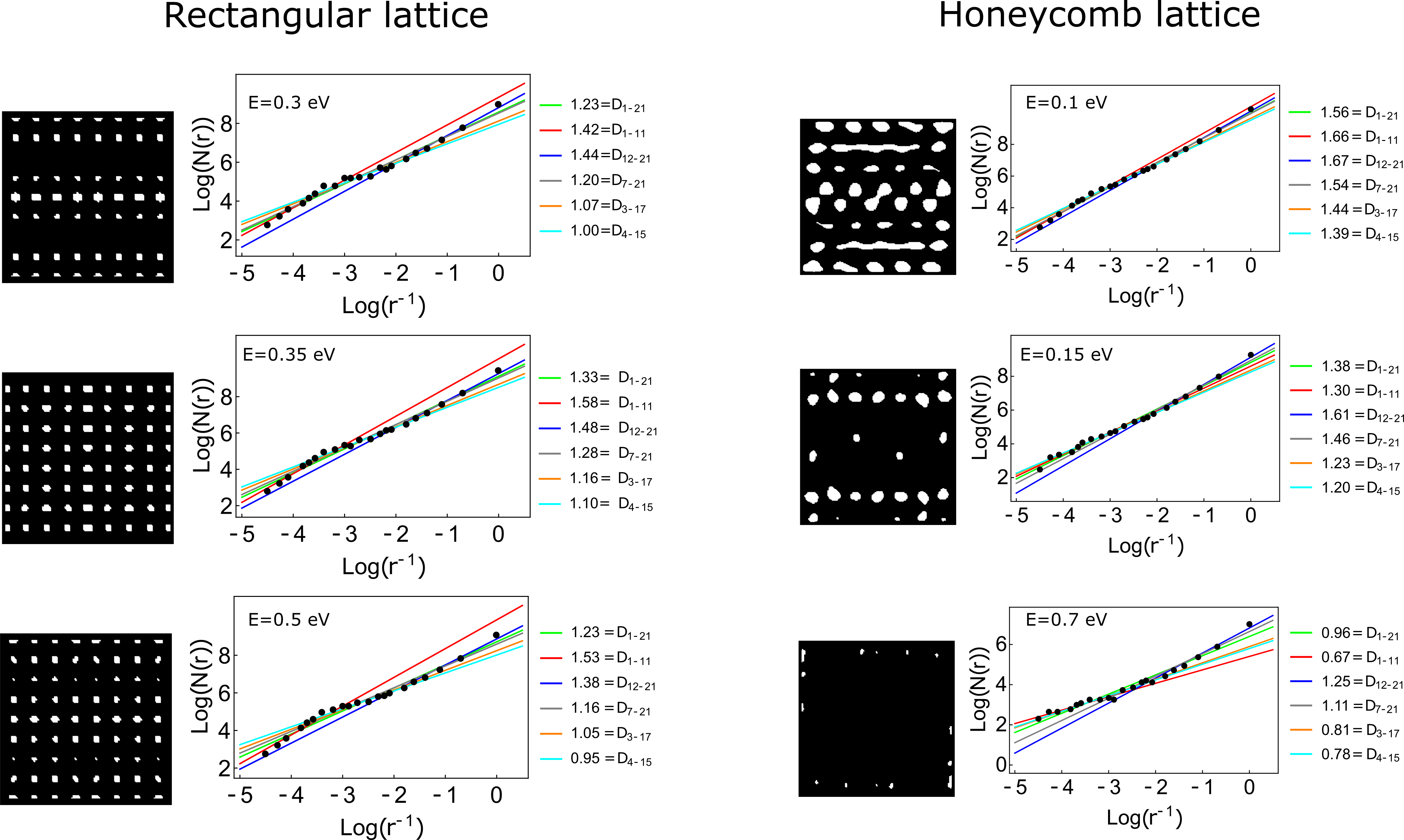}
\caption{This figure shows the binary images with a threshold of 75\% and the log-logplots for certain outliers in the dimension. For some energies, the dimensions of the rectangular and honeycomb lattices deviate significantly from 2. It is clear from the binary images that the threshold is set too high.}
\label{figoutliers}
\end{figure}
Some comments are in order regarding the interpretation of the dimension. First, we consider a $G(3)$ \Spi triangle with a finite thickness instead of an infinite generation or mathematical fractal. This implies that only a relatively small subset of box sizes should be used in the box-counting analysis. Therefore, the slope is steeper when using small boxes and the dimension for the G3 \Spi triangle converges towards $D=2$. For a mathematical fractal, the smaller boxes will still measure holes in the image and the slope will deviate less from the $D=1.58$ slope, which can be observed in Fig.~\ref{figsquaretriangle}. Second, the images that we use correspond to finite lattices and consist of a fixed number of pixels (set by the interpolation and the experimental and numerical resolution). Due to the finite size, the holes in the image have a width of several pixels which decreases when the lattice is larger (ultimately resulting in an image in which the holes are (almost) not visible anymore for the non-fractal lattices). This restricts the possible length scales that can be considered in the analysis, causing deviations from $D=2$ for e.g. the rectangular lattice when using small box sizes. In principle, infinitely large (rectangular) lattices will show a $D=2$ for a large number of box sizes. Given the finite size of the structures, small deviations in the LDOS and the presence of irregularly shaped holes in the binarized data lead to a lower box-counting dimension. Third, we note that the threshold of 75\% is too high for certain images, such as the one at $E=-0.1\,$eV for the muffin-tin \Spi triangle in Fig.~\ref{threshold}, where the image is nearly empty. The dimension of these images drops below one because the boxes measure unevenly distributed points in the confining box. To further increase the accuracy of the box-counting dimension estimate, one should determine a separate threshold for each image.

Figure~\ref{fig4} shows the box-counting dimension as a function of energy for a G(1), G(2) and G(3) \Spi triangle, as well as for a rectangular and honeycomb lattice on the same square grid. The markers correspond to dimensions determined on images binarized using a threshold of 65\%. The indicated uncertainties correspond to the dimensions extracted from images binarized with threshold values of 55\% and 75\%.
When using small boxes only, $i.e.$ the blue slopes, we observe that the box-counting dimensions for a $G(1)$ \Spi triangle at 55\% threshold (giving the highest values) are similar to the dimension of the filled triangle ($cf.$ Fig.~\ref{figsquaretriangle}). These dimensions are smaller for the $G(2)$ and $G(3)$ triangle. This is a clear indication that the holes in the $G(2)$ and $G(3)$ triangle are measured by the smaller box sizes, which is a feature of a (self-similar) fractal structure. These effects are less prominent for the larger boxes. When the guidelines from Ref.~[\!\!\citenum{Foroutan2}] are followed, larger boxes should be used (red slopes) and we find dimensions close to that of a mathematical \Spi fractal. Further, we observe that the dimension around certain energies such as $E=-0.1\,$eV is lower than the dimension at other energies. This difference is most likely caused by the LDOS map featuring either mostly connected (bonding) or disconnected (non-bonding) features. A disconnected image requires a similar number of large boxes to cover the fractal set, but a smaller number of small boxes, reducing the slope and thus the fractal dimension. For the rectangular and honeycomb lattice, the box-counting dimension deviates significantly from 2 for certain energies. Upon examining the images in Fig.~\ref{figoutliers}, it is clear that for these images the thresholds are chosen too high. This is consistent with the previous observation that ideally the thresholds are set for each image individually.

\subsubsection{The influence of the mask}
\begin{figure}[h!]
\centering
\includegraphics[width=1.0\columnwidth]{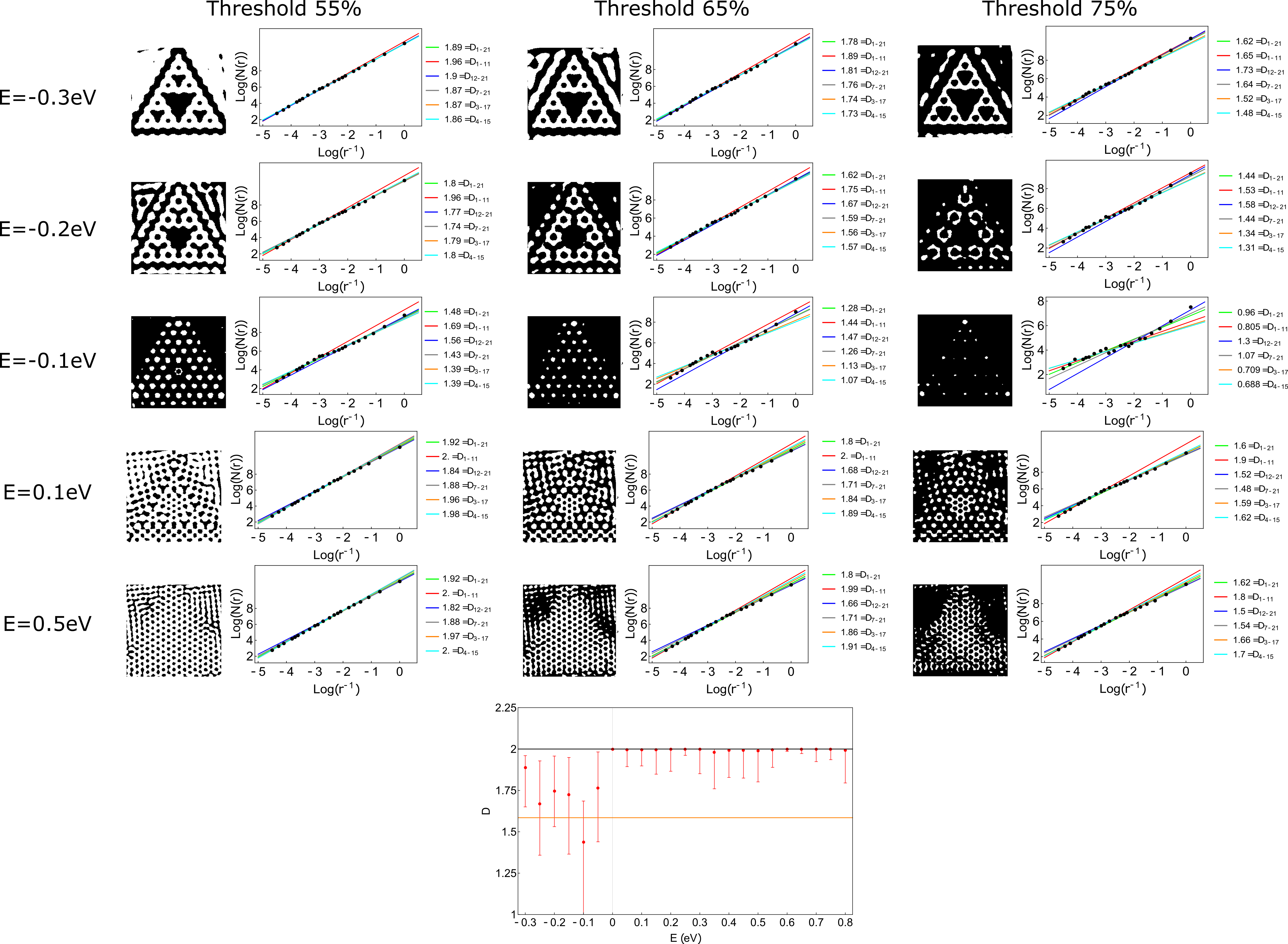}
\caption{The box-counting dimension without masks for the $G(3)$ \Spi fractal obtained from the muffin-tin method. The deviations for the different thresholds are large, which is due to the contribution of the standing waves surrounding the fractal. We observe that the influence of a mask is less important for energies $E<0\,$eV, which can be seen in the panel below the images, in which the dimension is shown for the red slopes. For energies $E>0\,$eV, the dimension of the LDOS of the \Spi triangle including background approaches 2, arising from the additional contribution of the emerging electron density between the CO molecules in the center of the \Spi triangle.}
\label{figurenomasks}
\end{figure}

When preparing the images for the calculation of the fractal dimension, we mask the background that we do not consider part of the structure. This includes masking the LDOS outside the triangle, but also the LDOS between the closely-packed CO molecules in the centers of the \Spi $G(2)$ and $G(3)$ triangles. Due to experimental limitations, it is not always possible to fully 'block' certain areas using the CO/Cu(111) platform. The masks serve as a proxy for the areas that should be excluded from the future experiments using other platforms (e.g. by etching or gating those areas).

\begin{figure}[h!]
\centering
\includegraphics[width=1.0\columnwidth]{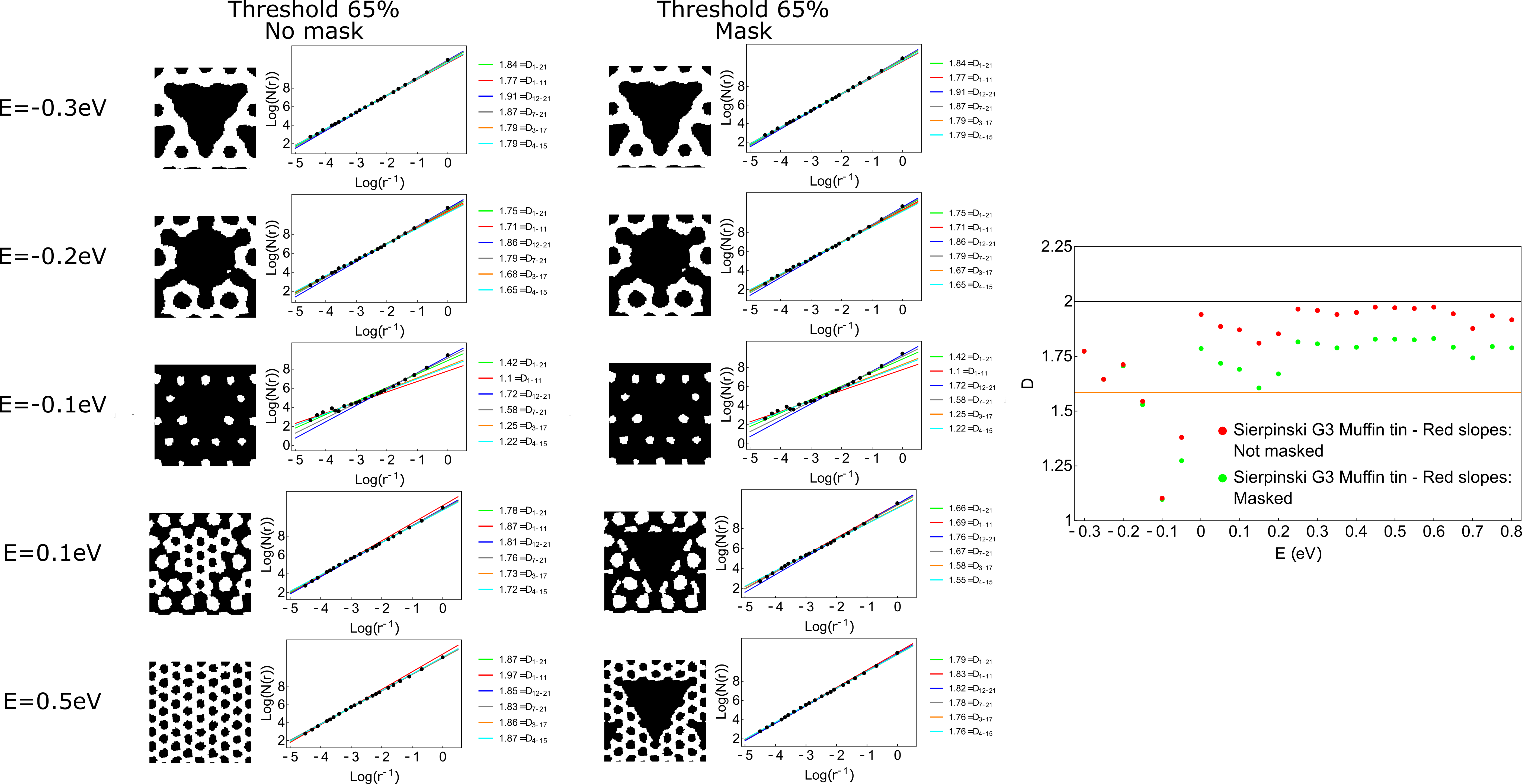}
\caption{The box-counting dimension without and with masks for the $G(3)$ generation \Spi fractal obtained from the muffin-tin method for a square subset of the images (a square is taken in the center of the $G(3)$ \Spi). Again, we observe that the influence of a mask is not significant for energies $E<0\,$eV, which can be seen in the panel in the right, where the dimension is shown for the red slopes. For energies $E > 0\,$eV, the dimension of the LDOS of the \Spi triangle including background is close to 2.}
\label{figmasksquares}
\end{figure}
In the following, we establish the influence of the masks by analyzing muffin-tin images of the $G(3)$ \Spi triangle without masks. The results are shown in Fig.~\ref{figurenomasks}. At energies below $0\,$eV, the box-counting dimensions of images with and without masks are similar (though slightly higher in comparison with Fig.~\ref{fig4}), and well below 2.0 for a threshold of 65\%. Hence, for these energies the masks have less influence. This means that at low energies, the electrons are confined to a fractal geometry. At higher energies, a more pronounced LDOS emerges in-between the closely-packed CO molecules and outside the lattice. Consequently, for energies above $0\,$eV, the box-counting analysis of the raw images without masks approaches a value of 2.0 (red slopes). Further, as can be seen in the images for $E=0.5\,$eV, the intensity in the LDOS at the artificial-atomic sites at high energies can no longer be described by $s$-wavefunctions, and $p$-bands should be taken into account~\cite{Slot2019}. As a result, the LDOS displays maxima in between the artificial $s$-wave \Spi atoms and the wavefunction map can be thought of as a honeycomb lattice that has a dimension of 2 (red slopes).
\\
To elucidate the influence of the masks further, we performed a box-counting analysis on a square subset of the \Spi G3 triangle at a threshold of 65\%, see Fig.~\ref{figmasksquares}. The results are consistent with those obtained for the full \Spi triangle on a square grid. The masks do not significantly influence the analysis at lower energies with a threshold of 65\%, whereas for higher energies the box-counting dimension is larger than for masked images.

\begin{figure}[h!]
\centering
\includegraphics[width=1.0\columnwidth]{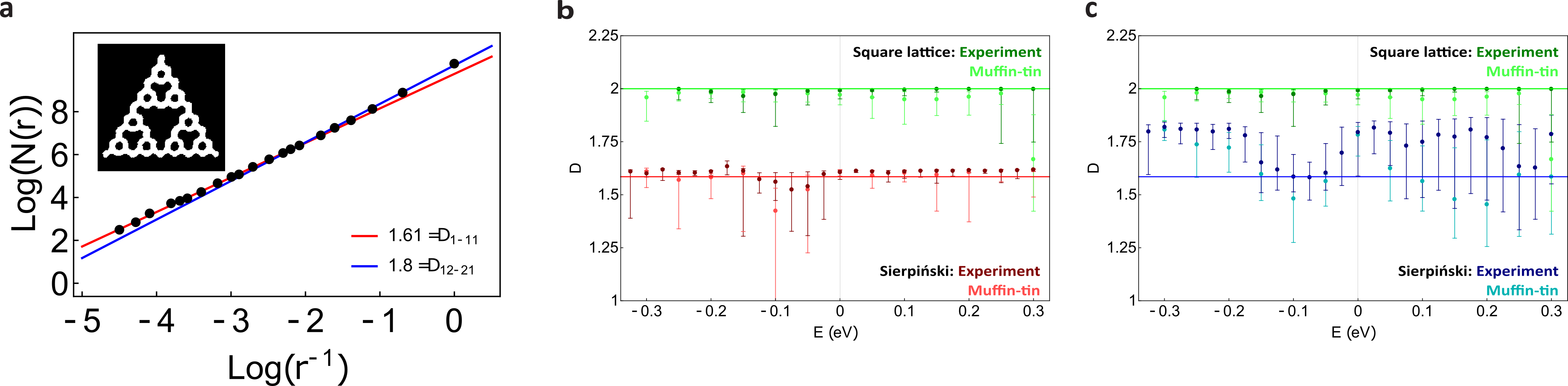}
\caption{\textbf{Fractal dimension of the \Spi wavefunction maps when the triangle is positioned at the top of the box.} \textbf{a}, The box-counting dimension of the wavefunction map acquired at $V = -0.325 \,$V is obtained from the slope of $\log(N)$ vs. $\log(r^{-1})$ where either the first 11 datapoints starting from the left (largest boxes, red slope) or the last 10 datapoints (smallest boxes, blue slope) are taken into account. Inset: binary map of the experimental \Spi LDOS map with a threshold of 65\%. \textbf{b}, Determination of the fractal dimensions of the LDOS of the $G(3)$ \Spi triangle (red) using the 11 largest boxes (red slopes) and comparison with the 2D rectangular lattice from Ref.~[\!\!\citenum{Slot20172}] (green) for the experimental (dark) and muffin-tin (light) wavefunction maps. The solid lines indicate the geometric \Spi Hausdorff dimension ($D = 1.58$) and that of the rectangular lattice ($D=2$). The error bars display the error in determining the fractal dimension at different LDOS thresholds, with thresholds of 55\% and 75\% giving the top and bottom of the error bar, respectively. \textbf{c}, The corresponding dimensions for the blue slopes. The main differences when placing the triangle at the top of the box are some small deviations in the results for the red slopes in comparison with the red slopes in Fig.~3 in the main text.}
\label{Fig1}
\end{figure}

\subsubsection{Minimum box-counting and skeleton analysis}
Here, we look into two effective approaches to improve the box-counting method for the $G(3)$ \Spi images from the muffin-tin calculation: minimum box-counting and an analysis of the skeleton images~\cite{Foroutan2}. The exact vertical position of the triangle on the square grid slightly influences the outcome of the box-counting analysis, see Fig.~\ref{Fig1}. The resulting differences in the dimension are mainly visible for the large box sizes (red slopes), but they remain small. The minimum box-counting procedure makes sure that the minimum number of boxes that is needed to cover the image is used, which provides a more accurate estimation of the dimension. Further, skeleton images of a 2D image provide a better determination of the slope since the deviation for smaller box sizes tends towards a slope of 1 for a line instead of 2 for solid images. In this way, it is easier to visualize where the points start to deviate from the straight line, and the slope can be determined more accurately. 

\begin{figure}[!h]
\centering
\includegraphics[width=1.0\textwidth]{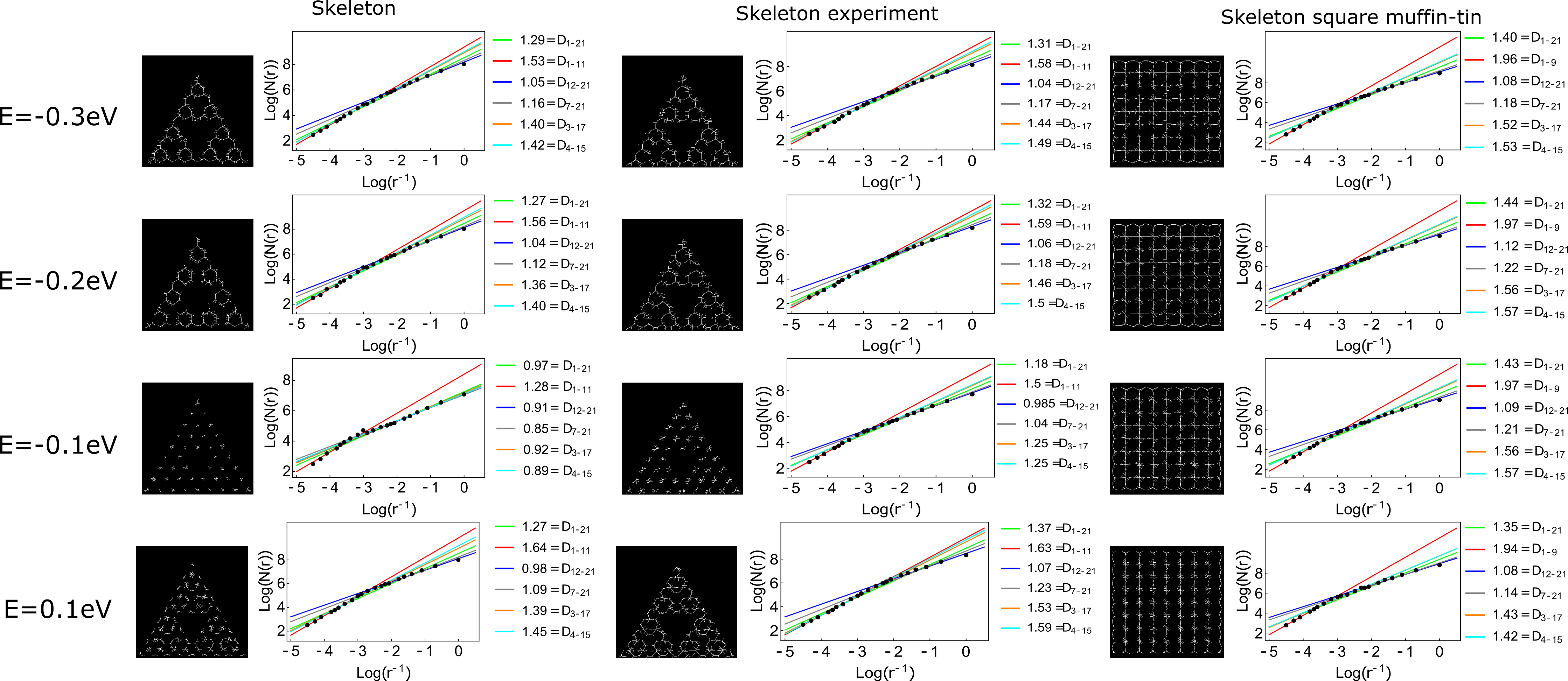}
\caption{Results for the dimension of the skeletonized muffin-tin wavefunction maps with a threshold of 65\%. The slope changes approximately after the 11th datapoint for the \Spi and experimental 'square' (rectangular) lattice (not shown), and for the theoretical rectangular lattice after the 9th datapoint starting from the left. The slope found for the skeletonized images is slightly lower than the one found for the images shown before, since they better capture the fractal nature of the structure. }
\label{fig5}
\end{figure}

\begin{figure}[!h]
\centering
\includegraphics[width=1.0\textwidth]{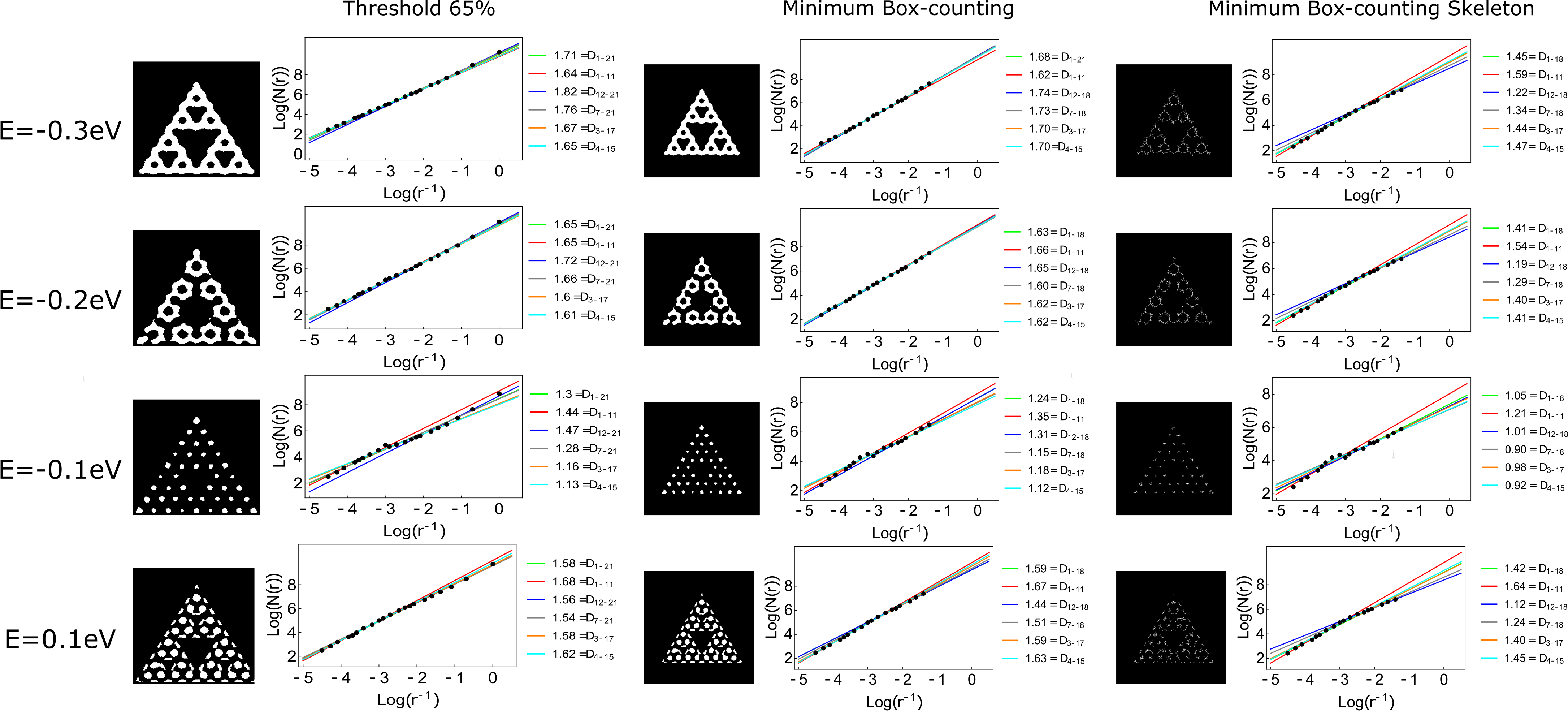}
\caption{Results for the minimum box-counting as mentioned in the text for the muffin-tin wavefunction maps with a binarization threshold of 65\%. The slopes deviate slightly from the results obtained without the minimum box-counting (shown on the left). Note that the box sizes 1-3 are not present in the minimum box-counting method.}
\label{fig6}
\end{figure}

\begin{figure}[!h]
\centering
\includegraphics[width=1.0\textwidth]{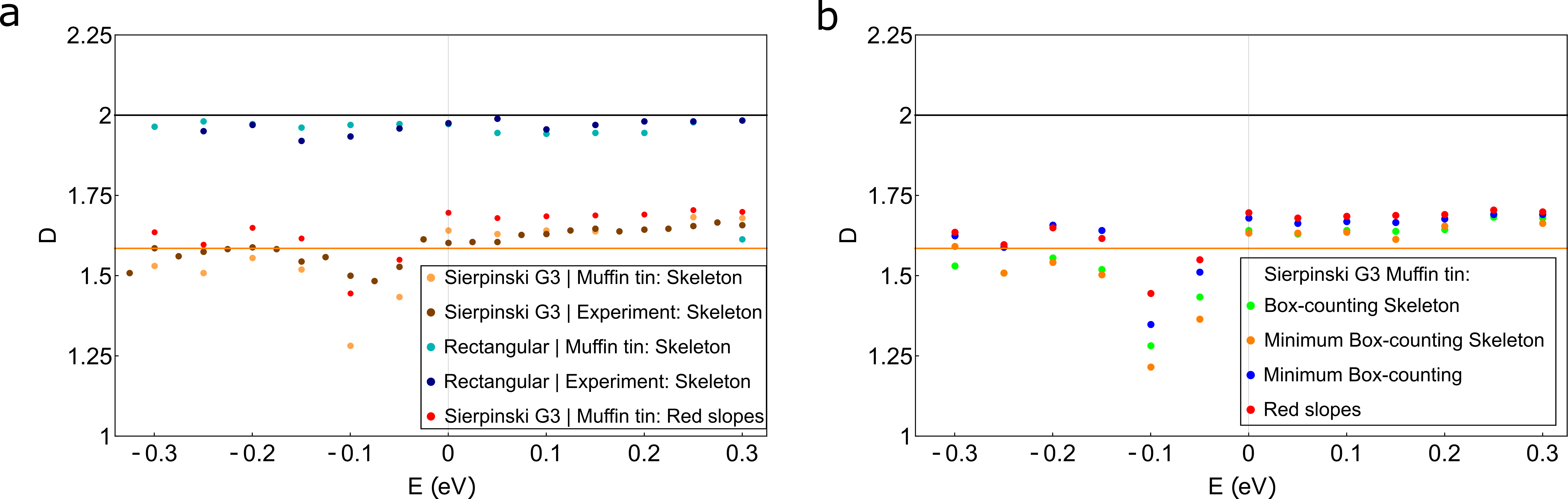}
\caption{The dimensions obtained with \textbf{(a)} the skeletonized images and  \textbf{(b)} with the minimum box-counting in comparison with the box-counting shown in previous sections. }
\label{fig7}
\end{figure}

For the minimum box-counting, we place the image with 65\% binarization threshold of the same size as before on a $500\,$x$500\,$ grid and position it on 100 different grid positions. The grid positions cover a range of 90 pixels in $x$ and $y$ direction, with a step size of 10 pixels. For a given box size, the minimum number of boxes that is needed to cover the image is determined and used in the calculation of the fractal dimension. In this calculation, the three smallest box sizes (1-3 pixels) are not included. 
\\ To create the skeleton images, we first select a threshold of 65\% and subsequently skeletonize the image by using the standard function \textit{SkeletonTransform} in Mathematica 12.0. The results for the minimum box-counting and the skeletonized images are shown in Figs.~\ref{fig5}-\ref{fig7}. We observe that the results are always close to the ones obtained for the red slopes reported in the previous sections.

\subsubsection{Scaling analysis}

\begin{figure}[!h]
\centering
\includegraphics[width=1.0\textwidth]{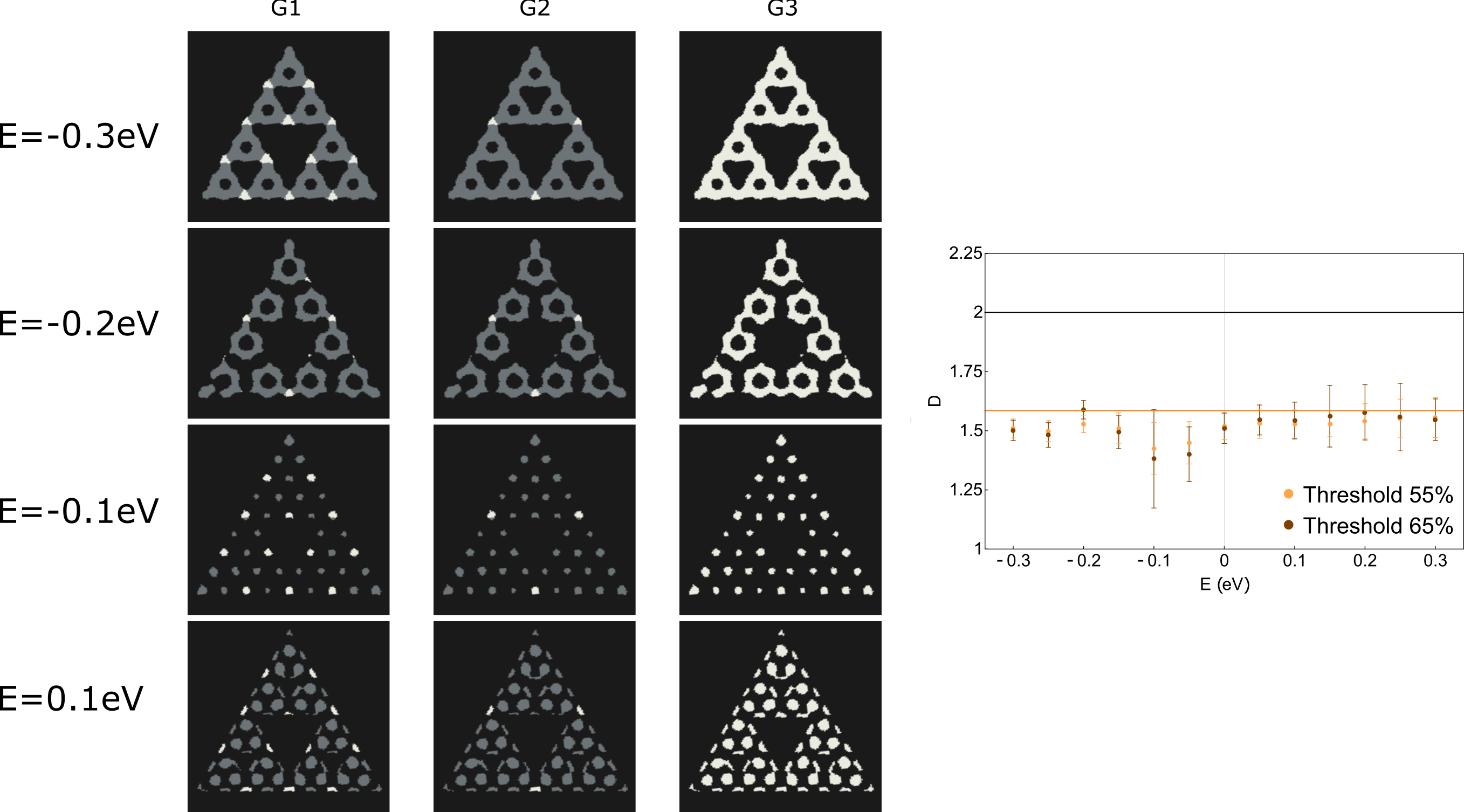}
\caption{The dimension obtained within the scaling analysis for the muffin-tin wavefunction map. In the left panel, the different subsets are shown for $G(1)$ and $G(2)$ in the $G(3)$ \Spi triangle. The grey part in $G(1)$ and $G(2)$ is part of a single subset, whereas the white part contains an overlap between two subsets. The scaling dimension is shown on the right for the thresholds $55$\% and $65$\%, where the error bar is the standard deviation. We observe that the dimension is close to 1.58 for the \Spi triangle, and differs from $D=2$ that is calculated for a solid object. }
\label{scalingMT1}
\end{figure}

\begin{figure}[!h]
\centering
\includegraphics[width=1.0\textwidth]{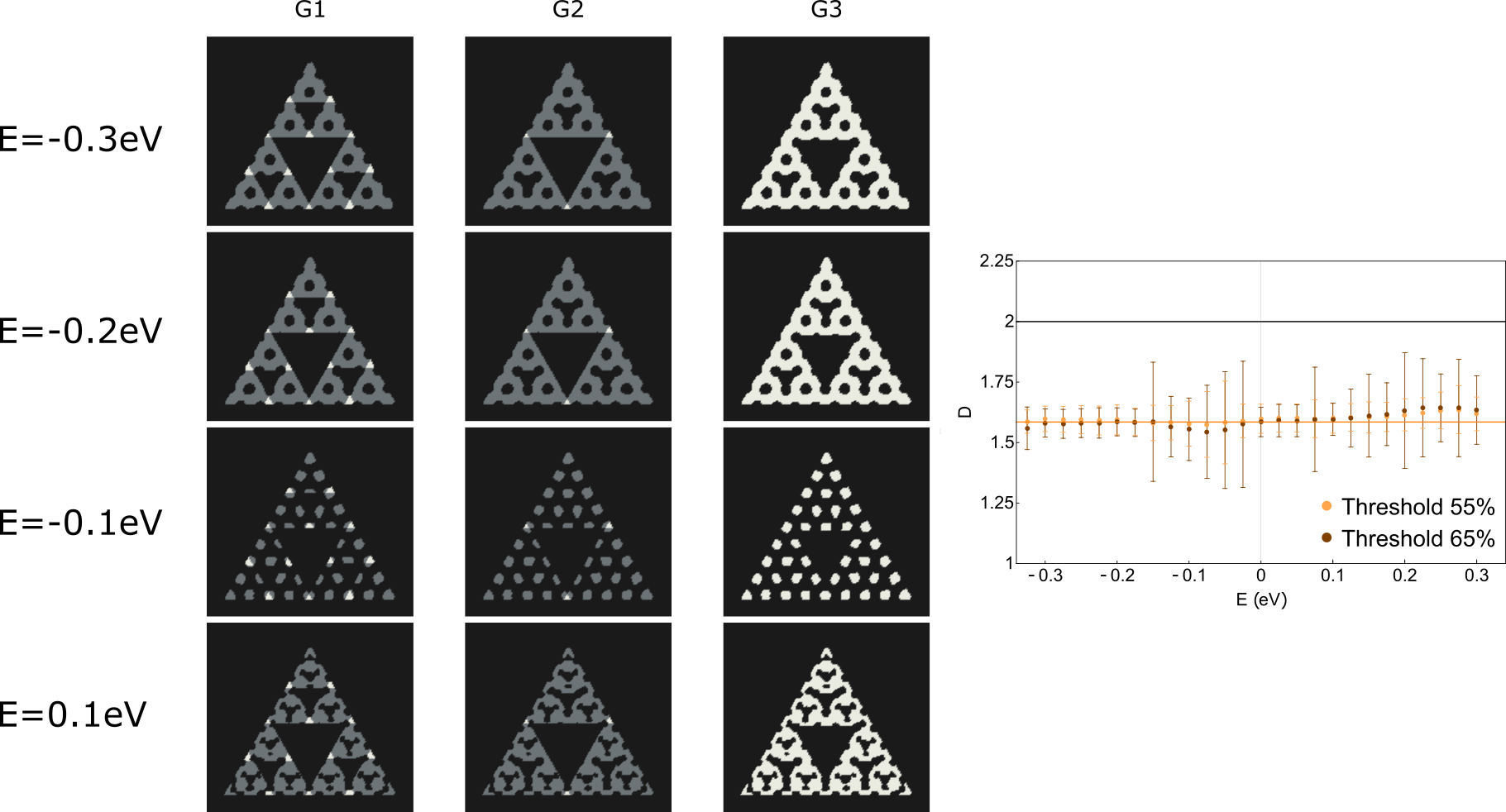}
\caption{The dimension obtained within the scaling analysis for the experimental maps. Again, we observe that the dimension is close to $D=1.58$ of the \Spi triangle.  }
\label{scalingExp}
\end{figure}

\begin{figure}[!h]
\centering
\includegraphics[width=1.0\textwidth]{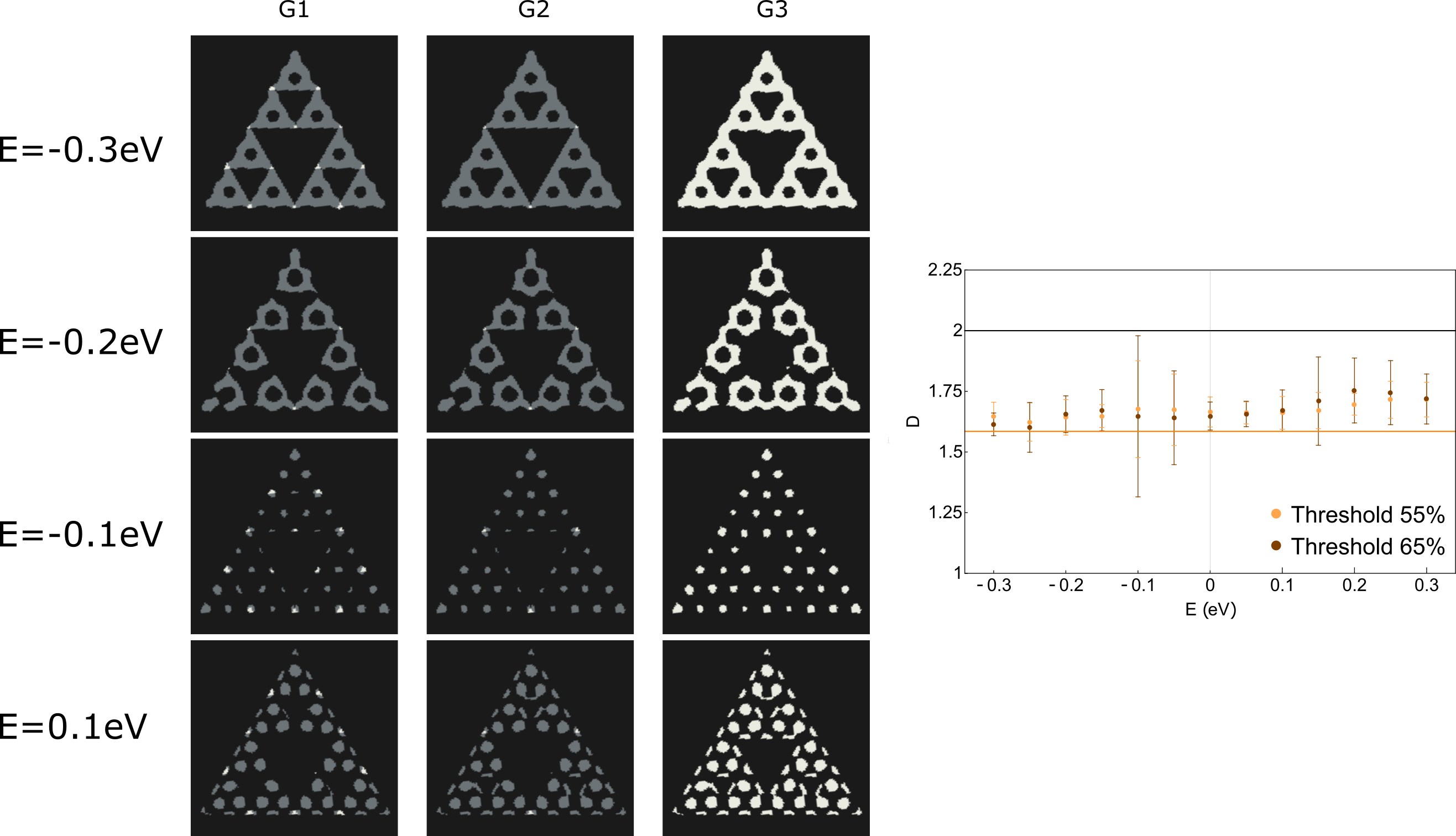}
\caption{The dimension obtained within the scaling analysis where the subsets are chosen to be smaller in comparison with Fig.~\ref{scalingMT1} (the white areas are chosen smaller and there is almost no overlap between the subsets). In this case, the resulting dimension is slightly above 1.58 and still well below $D = 2$ for the solid triangle. }
\label{scalingMT2}
\end{figure}

In this section, we consider a scaling analysis to calculate the fractal dimension of the third generation \Spi. The scaling of an object is defined as $N=S^D$, where $N$ is the number of measuring elements (such as sticks or number of pixels), $S$ the scaling factor, and $D$ the dimension. In our case, the number of pixels in a certain area is used as the measuring element to obtain the fractal dimension of the third generation \Spi fractals. By construction, the $G(3)$ \Spi fractal is composed of three $G(2)$ \Spi fractal subsets related by a scaling factor $S = 2$. In their turn, the $G(2)$ \Spi fractals are composed of three $G(1)$ \Spi fractal subsets, also related by a scaling factor $S = 2$. Hence, we can determine the number of pixels in the $G(2)$ \Spi subset and calculate the scaling dimension when comparing these to the number of pixels in the $G(3)$ \Spi using the relation $N_\text{G3}=N_\text{G2} \cdot 2^D$. Further, we can calculate the dimension by comparing the number of pixels in $G(1)$ with the number of pixels in $G(2)$ with $N_\text{G2}=N_\text{G1} \cdot 2^D$, or by comparing it with the number of pixels in $G(3)$ with $N_\text{G3}=N_\text{G1} \cdot 4^D$, where the scaling $S=4$ in the latter case. In the $G(3)$ \Spi, we can identify 9 $G(1)$ \Spi, 3 $G(2)$ \Spi and 1 $G(3)$ \Spi subsets. Thus, to calculate the scaling dimension for $G(3)$ based on the $G(1)$ and $G(2)$ triangles of which it is composed, there are $3 + 9 + 27 = 39$ options ($G(2) \rightarrow G(3)$, $G(1) \rightarrow G(3)$, and $G(1) \rightarrow G(2) \rightarrow G(3)$, respectively). In the following, we estimate the scaling dimension based on the mean value of the result for these 39 possibilities. A binarization threshold of 55\% and 65\% is used for both the experimental and the muffin-tin images.
For a solid object such as a triangle, this analysis results in $D = 2$.\\
\\
The results for the muffin-tin and experimental wavefunction maps are shown in Fig.~\ref{scalingMT1} and~\ref{scalingExp}, respectively. The images show which parts were considered as $G(1)$ and $G(2)$ subsets of the $G(3)$ \Spi triangle with a threshold of 65\%. In these images, the grey part is part of a single subset, whereas the white part displays an overlap between two subsets (the pixels in the white part are counted in both subsets). The corresponding dimensions for a threshold of 55\% and 65\% are shown in the right panel, where the error bar indicates the standard deviation of the calculated dimensions. We observe that the scaling dimension is qualitatively similar to the box-counting dimension and that the scaling dimension is close to $D=1.58$. \\
\\
The considered $G(3)$ \Spi fractal has a finite thickness, which results in a choice in determining what is part of the $G(1)$ subset in a $G(3)$ \Spi fractal. Therefore, we show the dimension for a different choice for the MT maps in Fig.~\ref{scalingMT2}. It can be clearly seen that, even though the different choices result in slightly different values, the dimension is close to the one of a mathematical \Spi triangle for all energies.

\subsubsection{Concluding remarks on the box-counting dimension}
The various box-counting analyses and the scaling analysis qualitatively provide similar results and all show a dimension well below $D=2$ for the \Spi triangle and, in most cases, a dimension close to $D=1.58$. However, as described above, the analyses applied to finite-size lattices (as done here) is subject to limitations and subjective choices (which box sizes should be taken into account in the analysis, a physical G(3) \Spi triangle vs. a mathematical fractal, the finite size of the square and honeycomb lattice, which threshold is chosen, triangular vs. rectangular shapes). Consequently, one should rather focus on the trends found rather than on the absolute values of the reported dimensions.

\subsection{Fourier analysis of the LDOS}

The wavefunction maps display a standing wave pattern, originating from the interference of the electronic wavefunctions that are scattered by the CO molecules. An analysis of the electronic standing-wave patterns at different energies is a powerful tool to resolve the energy dispersion of the 2D electron gas \cite{Hasegawa, CrommieCu111}, which is confined to the \Spi geometry in our study. We will use a Fourier analysis of the LDOS maps (also coined Fourier-transform scanning tunnelling spectroscopy or "FT-STS") \cite{Petersen, Fujita, Simon} for the in-depth analysis of the standing-wave patterns. The \Spi triangle lacks translational symmetry, and therefore we will make use of the pseudo-Brillouin zone~\cite{Nicola2, Gambaudo} in the following discussion.
\begin{figure}[!h]
\centering
\includegraphics[width=1.0\textwidth]{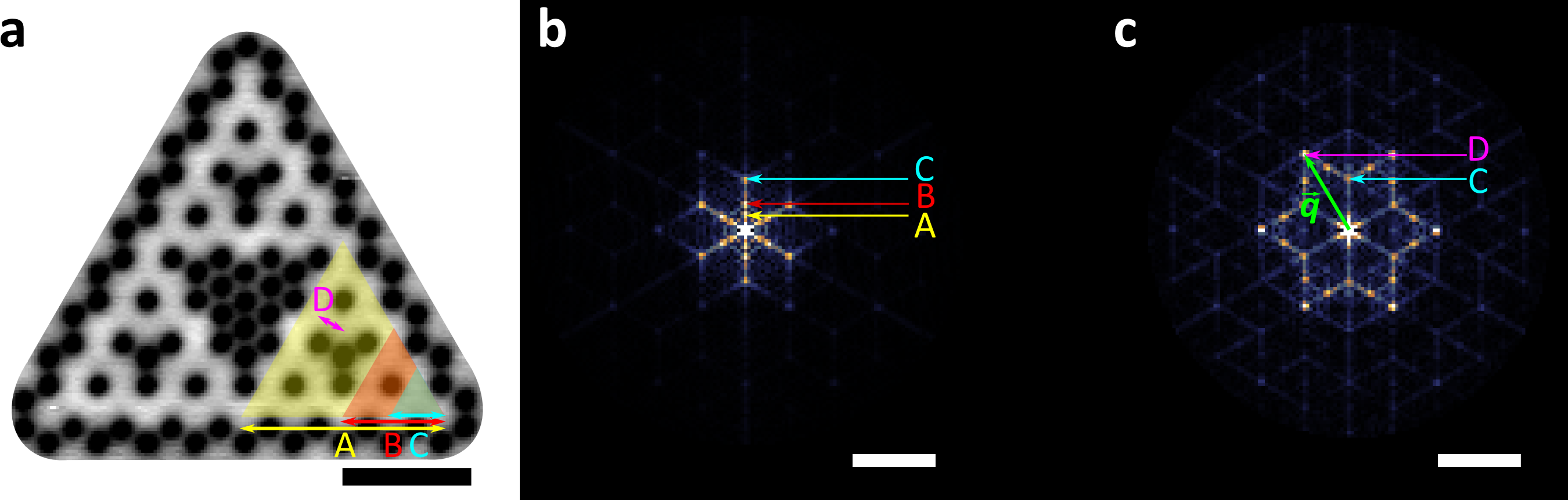}
\caption{\textbf{a}, Real-space experimental LDOS map at $V=-0.325\,$V, in which the length of a $G(2)$ side (\emph{A}), the length of a $G(1)$ side (\emph{B}), the NNN distance (\emph{C}), and the NN distance (\emph{D}) are indicated. Scale bar: $5\,$nm. \textbf{b}, Fast-Fourier transform of the experimental LDOS map at $V = -0.325\,$V shown in \textbf{a}. \textbf{c}, Fast-Fourier transform of the experimental LDOS map at $V = +0.100\,$V. The points in reciprocal space corresponding to the distances in \textbf{a} are indicated with arrows in \textbf{b} and \textbf{c}. An example of a scattering vector $\vec{q}$, pointing to \emph{D}, is indicated in green. Scale bar: $q = 6\,$nm$^{-1}$.  
}
\label{PeakAssignment}
\end{figure}
First, we relate several features in the obtained fast-Fourier transform (FFT) to the real-space wavelengths of the standing waves. Fig.~\ref{PeakAssignment} shows the real-space experimental wavefunction map at $V = -0.325\,$V with the relevant distances \emph{A}, \emph{B}, \emph{C}, and \emph{D} (\textbf{a}) along with the Fourier transform at $V = -0.325\,$V (\textbf{b}) and $V = 0.100\,$V (\textbf{c}). The magnitude of the scattering vector $q$ is related to the real-space distances $d$ of the \Spi structure as $q = \frac{2 \pi}{d} \cdot \frac{2}{\sqrt{3}}$, where the factor $\frac{2}{\sqrt{3}}$ originates from the triangular symmetry of the LDOS  converted to Fourier space. We thus obtain the following relevant distances and corresponding $q$ values:
\begin{center}
\begin{tabular}{ |l|c|c|c| } 
 \hline
 & $q$ [nm$^{-1}$]& $d$ [nm] & Definition \\ \hline
 A & 0.94 & 7.68 & Side length $G(2)$ \\ 
B & 1.89 & 3.84 & Side length $G(1)$ \\ 
C & 3.78 &  1.92 & NNN distance\\ 
D & 6.55 &  1.11 & NN distance \\ 
 \hline
\end{tabular}
\end{center}
These values were obtained from the geometry and experimentally observed in the FFTs of the LDOS maps, as indicated in Fig.~\ref{PeakAssignment}. Note that $q = 2 k$, as the LDOS is proportional to the norm square of the wavefunction \cite{Petersen,Hoermandinger}. For clarity, we used $k$ instead of $q$ in the main text.
Next, we point out that the self-similarity of the \Spi triangle in real space is directly reflected in momentum space, which is self-similar as well \cite{Lakhtakia_1986}. Each additional generation in real space adds a specific point to the Fourier transform. This was used in the main text to decompose the $G(3)$ to a $G(2)$ triangle of the same size by removal of the blue triangles in Fig.~\ref{PeakAssignment} (removal of the NNN-distance, C) and to decompose the $G(2)$ to a $G(1)$ triangle of the same size by removal of the red triangles (removal of the side length of the original $G(1)$, B). 
Finally, we study the energy $E$ vs. momentum $k$ relation of the \Spi electrons from the Fourier transforms at several energies. The standing waves around the \Spi triangle have been cropped, so only the electronic waves confined inside the \Spi geometry contribute. In Fig.~\ref{E_k}a-d, we recognize that the high-intensity maxima occur at higher $q$-values with increasing bias voltage. For instance, characteristic maxima at $-0.325\,$V, $-0.200\,$V, $-0.100\,$V, and $0.100\,$V  are positioned at $q = 1.2\,$nm$^{-1}$ \& $2.0 \,$nm$^{-1}$ (red), $q = 3.2\,$nm$^{-1}$ (green), $q = 3.8\,$nm$^{-1}$ (purple, NNN distances), and $q = 4.8\,$nm$^{-1}$ (orange), respectively (uncertainty $\pm$0.3 nm$^{-1}$). We note that for some energies, the maxima at neighboring discrete $q$-values are very pronounced and the intensity is affected by contributions of the triangular borders of the structure, due to which the $q$-values can be hard to assign (for example, both $q = 1.2\,$nm$^{-1}$ and $q = 2.0\,$nm$^{-1}$ are candidates for characteristic discrete values at $-0.325\,$V). The maxima at $q = 3.8\,$nm$^{-1}$ and $q = 6.4\,$nm$^{-1}$ are slightly visible at all energies, as they correspond to the NNN and NN distance between not only the electronic sites, but also between the CO molecules themselves. In Fig.~\ref{E_k}e, we plotted the above $E(k)$ values as well as intermediate values, where we used $q = 2k$. We observe that only discrete values of the momentum occur and the energy vs $k$ exhibits some jumps, demonstrating the confinement of the electrons, which form standing waves inside the \Spi geometry. A similar result has been observed for a checkerboard lattice in Ref.~[\!\!\citenum{Otte2}] and for cobalt islands in Ref.~[\!\!\citenum{Bode}]. Therefore, this is not a fractal feature, but can be attributed to the confinement of the electrons to the lattice, with the $k$-values related to the specific lattice geometry. Notably, the discrete $k$-values correspond to a small energy region rather than a discrete energy value, as the same standing waves are observed throughout this energy region. 
We observe that the $E(k)$ value is flatter than the free surface-state electron parabola with $E_0 = -0.45\,$eV and $m^* = 0.42 m_e$~\cite{Burgi2}, plotted in blue, due to the hybridization of the surface and bulk states. Again, while the statistical error is small, undesired contributions of the triangular borders of the structure lead to the risk of assigning the wrong discrete $k$ value to a certain energy. The quantitative result thus needs to be interpreted with caution.  

\begin{figure}[!h]
\centering
\includegraphics[width=0.55\textwidth]{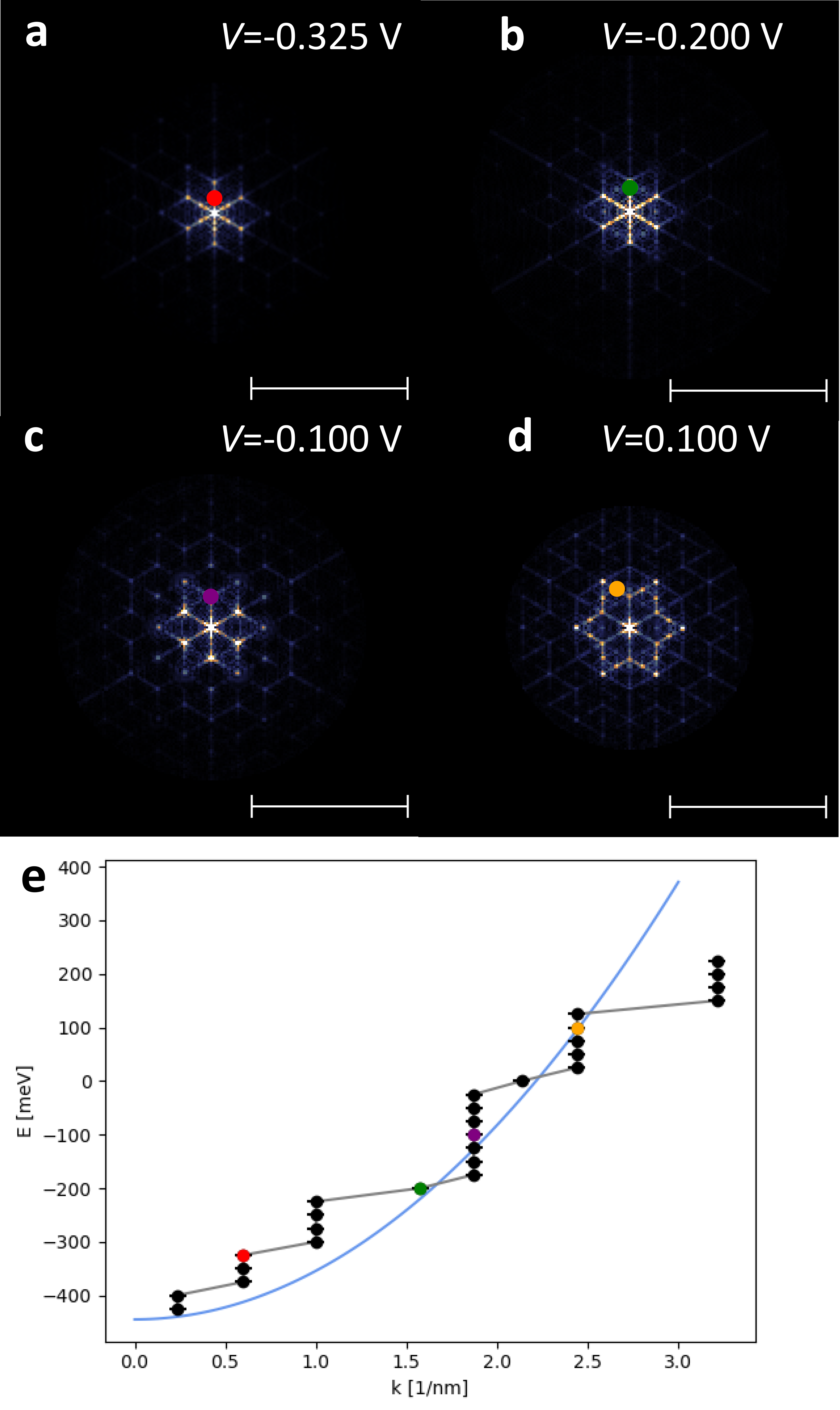}
\caption{\textbf{a-d}, Fourier transform of the differential conductance maps at $-0.325\,$V, $-0.200\,$V, $-0.100\,$V, and $0.100\,$V, respectively. The most pronounced $q$-values are indicated with a colored dot. Scale bar: $q = 9.4\,$nm$^{-1}.$ \textbf{e}, The characteristic $k$-values at their respective energies for the maps in \textbf{a-d} (colored dots) as well as for maps at intermediate energies. The blue curve indicates the free electron-like parabolic dispersion relation $E(k)$ for Cu(111) surface-state electrons.}
\label{E_k}
\end{figure}


\end{document}